# Ballistic to diffusive crossover of heat flow in graphene ribbons

Myung-Ho Bae[1,2,*], Zuanyi Li[1,3,*], Zlatan Aksamija[4], Pierre N. Martin[2], Feng Xiong[1,2,5], Zhun-Yong Ong[1,3], Irena Knezevic[4], and Eric Pop[1,2,5]

[1]*Micro & Nanotechnology Lab, Univ. Illinois, Urbana-Champaign, IL 61801, USA*
[2]*Dept. of Electrical & Computer Engineering, Univ. Illinois, Urbana-Champaign, IL 61801, USA*
[3]*Dept. of Physics, Univ. Illinois, Urbana-Champaign, IL 61801, USA*
[4]*Dept. of Electrical & Computer Engineering, Univ. Wisconsin-Madison, Madison, WI 53706, USA*
[5]*Beckman Institute, Univ. Illinois, Urbana-Champaign, IL 61801, USA*

[*]These authors contributed equally.

**Abstract:**

Heat flow in nanomaterials is an important area of study, with both fundamental and technological implications. However, little is known about heat flow in two-dimensional (2D) devices or interconnects with dimensions comparable to the phonon mean free path (mfp). Here, we find that short, quarter-micron graphene samples reach ~35% of the ballistic heat conductance limit up to room temperature, enabled by the relatively large phonon mfp (~100 nm) in substrate-supported graphene. In contrast, patterning similar samples into nanoribbons (GNRs) leads to a diffusive heat flow regime that is controlled by ribbon width and edge disorder. In the edge-controlled regime, the GNR thermal conductivity scales with width approximately as $\sim W^{1.8 \pm 0.3}$, being about 100 $Wm^{-1}K^{-1}$ in 65-nm-wide GNRs, at room temperature. Manipulation of device dimensions on the scale of the phonon mfp can be used to achieve full control of their heat-carrying properties, approaching fundamentally limited upper or lower bounds.

The thermal properties of graphene are derived from those of graphite, and are similarly anisotropic. The in-plane thermal conductivity of isolated graphene is high, ~2000 $Wm^{-1}K^{-1}$ at room temperature, due to the strong $sp^2$ bonding and relatively small mass of carbon atoms[1-3]. Heat flow in the cross-plane direction is nearly a thousand times weaker, limited by van der Waals interactions with the environment (for graphene)[4] or between graphene sheets (for graphite)[1, 2]. Recent studies have suggested that the thermal conductivity of graphene is altered when in contact with a substrate through the interaction between vibrational modes (phonons) of graphene and those of the substrate[5-8]. However an understanding of heat flow properties in nanometer scale samples of graphene [or any other two-dimensional (2D) materials] is currently lacking.

By comparison, most graphene studies have focused on its electrical properties when confined to scales on the order of the carrier mean free path (mfp)[9-14]. For example, these have found that "short" devices exhibit near-ballistic behavior[9], Fabry-Perot wave interference[12], and "narrow" nanoribbons display a steep reduction of charge carrier mobility[11, 13]. Previous studies do exist for heat flow in three-dimensional (3D) structures like nanowires and nanoscale films. For instance, ballistic heat flow was observed in suspended GaAs bridges[15] and silicon nitride membranes[16] at low temperatures, of the order 1 K. Conversely, suppression of thermal conductivity due to strong edge scattering effects was noted in narrow and rough silicon nanowires[17, 18], up to room temperature. Yet such effects have not been studied in 2D materials like graphene, and ballistic heat conduction has not been previously observed near room temperature in any material.

In this work we find that the thermal properties of graphene can be tuned in nanoscale devices comparable in size to the intrinsic phonon mfp. (By "intrinsic" thermal conductivity or phonon mfp we refer to that in large samples without edge effects, typically limited by phonon-phonon scattering in suspended graphene, and by substrate scattering in supported graphene, here $\lambda \approx 100$ nm at room temperature). The thermal conductance of "short" quarter-micron graphene reaches up to 35% of theoretical ballistic upper limits[19]. However, the thermal conductivity of "narrow" graphene nanoribbons (GNRs) is greatly reduced compared to that of "large" graphene samples. Importantly, we uncover that nanoengineering the GNR dimensions and edges is responsible for altering the effective phonon mfp, shifting heat flow from quasi-ballistic to diffusive regimes. These findings are highly relevant for all nanoscale graphene devices and interconnects, also suggesting new avenues to manipulate thermal transport in 2D and quasi-one-dimensional (1D) systems.

## Results

**Test structures and measurements.** Figure 1 illustrates several of our experimental test structures, showing graphene and GNR arrays supported on a $SiO_2$/Si substrate (see Methods and Supplementary Note 1). Long parallel metal lines serve as heater and thermometer sensors[5, 20], electrically insulated from the graphene by a thin $SiO_2$ layer. We perform heat flow measurements from 20 to 300 K on unpatterned graphene (Fig. 1a), control samples with the graphene etched off (Fig. 1b), and arrays of GNR widths $W \approx 130$, 85, 65 and 45 nm (Figs. 1c-d and Supplementary Fig. S2). Figure 1f shows the Raman spectra of representative samples, with no discernible D peak (no defects) in unpatterned graphene[4] and a D/G peak ratio of GNRs consistent with the presence of edge disorder[14, 21].

The measurement proceeds as follows. We pass a heating current through one metal line which sets up a temperature gradient across the sample, and we monitor changes in electrical resistance of the opposite electrode (see Methods and Supplementary Note 2). Both electrode resistances are calibrated over the full temperature range for each sample, allowing us to convert measured changes of resistance into changes of sensor temperature $\Delta T_S$ as a function of heater power $P_H$ (Supplementary Fig. S5). We also perform measurements after removing the exposed graphene with an oxygen plasma etch (Fig. 1b). This allows us to obtain the thermal properties of the parallel heat flow path through the contacts, supporting $SiO_2$ and substrate (Figs. S4 and S8). As a check on our method, we find the thermal conductivity of our $SiO_2$ layer in excellent agreement with well-known data from the literature (Supplementary Note 4 and Fig. S8) over the full temperature range. As a result of this exercise, we were also able to fit the thermal resistance of the $SiO_2$-Si interface (Supplementary Fig. S8c and eq. S1), generating one of the few available data sets on this quantity, to our knowledge.

To obtain the thermal properties of our samples we use 3D simulations of the structures with dimensions obtained from measurements by scanning electron microscopy (SEM) and atomic force microscopy (AFM), as shown in Figs. 1d-e and Supplementary Fig. S7. The model matches the measured and simulated $\Delta T_S$ and $P_H$, fitting the thermal conductance $G$ between heater and thermometer. The 3D simulations automatically include all known contact resistance effects, including those of the graphene-$SiO_2$ and $SiO_2$-metal interfaces, matched against data from the literature and our control experiments (Supplementary Note 3). To provide some simple estimates, the contact thermal resistance (per electrode width) is $R_C \approx 0.7$ m·KW$^{-1}$, the "wide" unpatterned graphene thermal resistance is $R_G \approx 2.5$ m·KW$^{-1}$, and that of the GNR arrays is in the range $R_{GNR} \approx 4$ to 32 m·KW$^{-1}$ (from widest to narrowest). The graphene is not patterned under the electrodes, thus the contact resistance remains the same for all samples. The 3D simulations also account for heat spreading through the underlying $SiO_2$, and our error bars include various uncertainties in all parameters (Supplementary Note 6).

Figure 2a displays in-plane thermal conductance per area ($G/A$) for our GNRs, for one of our unpatterned "short but wide" samples ($L \approx 260$ nm, $W \approx 12$ μm), and for the "large" sample ($L \approx 10$ μm) of Ref. [6]. Here $A$ is the cross-sectional area of heat flow, $A = WH$, where $W$ is the width and $H = 0.335$ nm is the thickness of the graphene samples. In parallel, Fig. 3 displays schematics of the size effects and the three transport regimes expected, corresponding to the samples measured in Fig. 2. Figure 2a also shows the theoretical ballistic thermal conductance of graphene[22-24], $G_{ball}/A$, calculated with the approach listed in Supplementary Note 9. By comparison, our "short" sample (schematic in Fig. 3b) has a thermal conductance ~35% of $G_{ball}/A$ at 200 K and ~30% at room temperature, indicating a regime of quasi-ballistic phonon transport (other similar samples are shown in Supplementary Fig. S9d). In contrast, the "large" sample from Ref. [6] (schematic in Fig. 3a) has a conductance per cross-sectional area <2% of the ballistic limit, being in the diffusive transport regime as expected ($W, L \gg \lambda$).

**Length dependence of thermal conductivity.** We recall that in the ballistic limit ($L \ll \lambda$) the conductance rather than the conductivity approaches a constant at a given temperature[22-24], $G_{ball}(T)$. Nevertheless, the thermal conductivity is the parameter typically used for calculating heat transport in practice, and for comparing different materials and systems. Thus, the well-known relationship $k = (G/A)L$ imposes the conductivity $k$ to become a function of length in the ballistic regime and to *decrease* as $L$ is reduced. This situation becomes evident when we plot the thermal conductivity in Fig. 2b, finding $k \approx 320$ Wm$^{-1}$K$^{-1}$ for our "short" and wide samples at room temperature (schematic Fig. 3b), almost a factor of two lower than the large graphene[6] (schematic Fig. 3a). We note that both unpatterned samples here and in Ref. [6] were supported by $SiO_2$, showed no discernible defects in the Raman spectra, and the measurements were repeated over three samples (Supplementary Note 5 and Fig. S9), with similar results obtained each time.

The transition of thermal conductivity from diffusive to ballistic can be captured through simple models[25], similar to the apparent mobility reduction during quasi-ballistic charge transport observed in short-channel transistors[26, 27]:

$$k(L) = \sum_{p}\left(\frac{A}{LG_{\mathrm{p,ball}}} + \frac{1}{k_{\mathrm{p,diff}}}\right)^{-1} \approx \frac{G_{\mathrm{ball}}}{A}\left[\frac{1}{L} + \frac{1}{(\pi/2)\lambda}\right]^{-1}. \qquad (1)$$

The first equality is a "three-color" model with $p$ the phonon mode (longitudinal acoustic, LA; transverse, TA; flexural, ZA), $G_{\mathrm{p,ball}}$ calculated using the appropriate dispersion[22], and $\sum k_{\mathrm{p,diff}} = k_{\mathrm{diff}}$ the diffusive thermal conductivity (~600 $\mathrm{Wm^{-1}K^{-1}}$ at 300 K)[6]. A simpler "gray" approximation can also be obtained by dropping the $p$ index, $k(L) \approx [A/(LG_{\mathrm{ball}}) + 1/k_{\mathrm{diff}}]^{-1}$, where $G_{\mathrm{ball}}/A \approx 4.2 \times 10^9$ $\mathrm{WK^{-1}m^{-2}}$ at room temperature[19] (see Supplement Note 9). The second expression in eq. 1 is a Landauer-like model[25, 28], with $\pi/2$ accounting for angle averaging[29] in 2D to obtain the phonon backscattering mfp. For convenience, we note that the ballistic thermal conductance of graphene can be approximated analytically as $G_{\mathrm{ball}}/A \approx [1/(4.4 \times 10^5\ T^{1.68}) + 1/(1.2 \times 10^{10})]^{-1}$ $\mathrm{WK^{-1}m^{-2}}$ over the temperature range 1-1000 K, as a fit to full numerical calculations (Supplementary Fig. S16).

We compare the simple models in eq. 1 with the experiments in Fig. 2c and find good agreement over a wide temperature range. The comparison also yields our first estimate of the intrinsic phonon mfp in $SiO_2$-supported graphene, $\lambda \approx (2/\pi)k_{\mathrm{diff}}/(G_{\mathrm{ball}}/A) \approx 90$ nm at 300 K and 115 nm at 150 K. (The same argument estimates an intrinsic phonon mfp $\lambda \approx 300$-600 nm in freely suspended graphene at 300 K, if a thermal conductivity 2000-4000 $\mathrm{Wm^{-1}K^{-1}}$ is used[1-3].) This phonon mfp is the key length scale which determines when the thermal conductivity of a sample becomes a function of its dimensions, in other words when $L$ and $W$ are comparable to $\lambda$. Based on Fig. 2c, we note that ballistic heat flow effects should become non-negligible in all $SiO_2$-supported graphene devices shorter than approximately 1 μm.

**Width dependence of thermal conductivity.** We now turn to the width-dependence of heat flow in narrow GNRs. Our experimental data in Figs. 2b and 2d show a clear decrease of thermal conductivity as the width $W$ is reduced to a regime comparable to the intrinsic phonon mfp. For instance, at room temperature $k \approx 230$, 170, 100, and 80 $\mathrm{Wm^{-1}K^{-1}}$ for GNRs of $W \approx 130$, 85, 65 and 45 nm, respectively, and same $L \approx 260$ nm. To understand this trend, we consider $k$ limited by phonon scattering with edge disorder[30, 31] through a simple empirical model with a functional form suggested by previous work on rough nanowires[32, 33] and GNR mobility[34]:

$$k_{\text{eff}}\left(W,L\right)\approx\left[\frac{1}{c}\left(\frac{\Delta}{W}\right)^{n}+\frac{1}{k(L)}\right]^{-1}. \qquad (2)$$

Here $\Delta$ is the root-mean-square (rms) edge roughness and $k(L)$ is given by eq. 1. The solid lines in Fig. 2d show good agreement with our GNR data ($L \approx 260$ nm) using $\Delta = 0.6$ nm and a best-fit exponent $n = 1.8 \pm 0.3$. The parameter $c = 0.04$ $\text{Wm}^{-1}\text{K}^{-1}$ can be used to fit the room-temperature data set and additional fitting discussion is provided in the Supplementary Note 9. (Note that we cannot assign overly great physical meaning to the parameter $c$ because the empirical model can only fit $\Delta^n/c$, not $\Delta$ or $c$ independently.) The simple model appears to be a good approximation in a regime with $\Delta \ll W$, where the data presented here were fitted. However it is likely that this simple functional dependence would change in a situation with extreme edge roughness[18], where the roughness correlation length (which cannot be directly quantified here) could also play an important role.

Nevertheless, the nearly $W$-squared dependence of thermal conductivity in narrow GNRs with edge roughness is consistent with previous findings for rough nanowires[32, 33], and also similar to that suggested by theoretical studies of GNR electron mobility[34]. The precise scaling with $\Delta$ is ostensibly more complex[30, 31] than can be captured in a simple model, as it depends on details of the phonon dispersion, the phonon wave vector, and indirectly on temperature. However the $\Delta$ estimated from the simple model presented above is similar to that from extensive numerical simulations below, and to that measured by transmission electron microscopy (TEM) on GNRs prepared under similar conditions[35]. Thus, the simple expressions given above can be taken as a practical model for heat flow in substrate-supported GNRs with edge roughness ($\Delta \ll W$) over a wide range of dimensions, corresponding to all size regimes in Fig. 3.

## Discussion

We consider the effects of measurement contacts and how they relate to the interpretation of sample length in the quasi-ballistic heat flow regime. As in studies of quasi-ballistic electrical transport[26, 27], we defined the "channel length" $L$ as the inside edge-to-edge distance between the heater and thermometer electrodes (Fig. 3c). Simple ballistic theory assumes contacts with an infinite number of modes, and instant thermalization of phonons at the edges of the contacts. The former is well approximated here by electrodes two hundred times thicker than the graphene sheet, however phonons may travel some distance below the contacts before equilibrating. The classical, continuum analog of this aspect is represented by the thermal transfer length ($L_T$) of heat flow from the graphene into the contacts[36], which is automatically accounted in our 3D simulations (Fig. 1e). However, a sub-continuum perspective[37] reveals that phonons only thermally equilibrate after traveling one mfp into the graphene under the contacts. Previous measurements of oxide-encased graphene[5] had estimated a thermal conductivity $k_{enc}$ = 50-100 $Wm^{-1}K^{-1}$, which suggests a phonon mfp $\lambda_{enc} = (2k_{enc}/\pi)/(G_{ball}/A) \approx$ 8-15 nm under the contacts. This adds at most 12% to our assumption of edge-to-edge sample length (here $L \approx 260$ nm), a small uncertainty which is comparable to the sample-to-sample variation from fabrication, and to the size of the symbols in Fig. 2c. (The relatively low thermal conductivity of encased monolayer graphene[5] is due to scattering with the $SiO_2$ sandwich, although some graphene damage from the $SiO_2$ evaporation[38] on top is also possible.)

To gain deeper insight into our experimental results, we employ a numerical solution of the Boltzmann transport equation (BTE) with a complete phonon dispersion[31, 39]. Our approach is similar to previous work[6, 40], but accounting for quasi-ballistic phonon propagation and edge disorder scattering in short and narrow GNRs, respectively (see Methods and Supplementary Notes 7 and 8). Figure 4a finds good agreement of thermal conductivity between our measurements and the BTE model across all samples and temperatures. We obtained the best fit for GNRs of width 130 and 85 nm with rms edge roughness $\Delta$ = 0.25 and 0.3 nm, where the gray bands in Fig. 4a correspond to ±5% variation around these values. For GNRs of widths 65 and 45 nm the gray bands correspond to edge roughness ranges $\Delta$ = 0.35-0.5 and 0.5-1 nm, respectively. We note that, unlike the empirical model of eq. 2, the best-fit BTE simulations do not use a unique value of edge roughness $\Delta$. This could indicate some natural sample-to-sample variation in edge roughness from the fabrication conditions, but it could also be due to certain edge scattering physics (such as edge roughness correlation[18] and phonon localization[41]) which are not yet captured by the BTE model.

Figure 4b examines the scaling of mfps by phonon mode, finding they are strongly reduced as the GNR width decreases below ~200 nm, similar to the thermal conductivity in Fig. 2d. The mfp for each phonon mode is calculated as an average over the entire frequency spectrum, weighted by the frequency-dependent heat capacity and group velocity (Supplementary eq. S19). We note that LA and TA modes, which have larger intrinsic mfps, are more strongly affected by GNR edge disorder. On the other hand ZA modes are predominantly limited by substrate scattering and consequently suffer less from edge disorder, consistently with recent findings from molecular dynamics (MD) simulations[7, 8].

Increasing edge disorder reduces phonon mfps (Supplementary Fig. S15d), and the thermal conductivity is expected to scale as shown in Fig. 4c. In the BTE model the edge roughness scattering is captured using a momentum-dependent specularity parameter (Supplementary eq. S11), meaning that small wavelength (large $q$) phonons are more strongly affected by line edge roughness. However, as $\Delta$ increases the specularity parameter saturates, marking a transition to fully diffuse edge scattering, and also to a regime where substrate scattering begins to dominate long wavelength phonons in substrate-supported samples. This transition cannot be captured by the simplified $\Delta^n$ dependence in the empirical model of eq. 2.

To further illustrate such distinctions, Fig. 4d displays the energy (frequency $\omega$) dependence of phonon mfps for a “small” GNR and a “large” $SiO_2$-supported graphene sample (corresponding to Figs. 3c and 3a, respectively). Low-frequency substrate scattering (proportional to $\sim 1/\omega^2$) dominates the large sample[6, 7], while scattering with edge disorder affects phonons with wavelengths comparable to, or smaller than, the roughness $\Delta$ (see Supplementary Note 7). Therefore, larger $\Delta$ can affect more long wavelength (low energy) phonons, but only up to $\Delta \sim 1$ nm, where the effect of the substrate begins to dominate in the long wavelength region. (Also seen in Fig. 4c.) Such a separation of frequency ranges affected by substrate and edge scattering could provide an interesting opportunity to tune both the total value and the spectral components of thermal transport in GNRs, by controlling the substrate and edge roughness independently.

It is instructive to examine some similarities and differences between our findings here vs. previous results regarding size effects on charge carrier mobility in GNRs with dimensions comparable to the phonon or electron mfp. The edge-limited thermal conductivity begins to fall off in GNRs narrower than approximately ~200 nm (Fig. 2d), or twice the phonon mfp. A similar trend was noted for the electrical mobility in GNRs[11], but with a fall-off at widths narrower than ~40 nm (Supplementary Fig. S11). These observations are consistent with the electron mfp being several times shorter[13, 42] than the phonon mfp in $SiO_2$-supported graphene, i.e. approximately 20 nm for the electron mfp vs. nearly ~100 nm for the phonon mfp at room temperature. Thus, edge disorder affects thermal transport more strongly than charge transport in GNRs of an intermedi-ate width (40 < $W$ < 200 nm), an effect that could be used to manipulate charge and heat flow independently in such nanostructures.

In conclusion, we investigated heat flow in $SiO_2$-supported graphene samples of dimensions comparable to the phonon mfp. Short devices ($L \sim \lambda$, corresponding to Fig. 3b schematic) have thermal conductance much higher than previously found in micron-sized samples, reaching 35% of the ballistic limit at 200 K and 30% (~1.2 GW $K^{-1}$ $m^{-2}$ ) at room temperature. However, narrow ribbons ($W \sim \lambda$, corresponding to Fig. 3c schematic) show decreased thermal conductivity due to phonon scattering with edge disorder. Thus, the usual meaning of thermal conductivity must be carefully interpreted when it becomes a function of sample dimensions. The results also suggest powerful means to tune heat flow in 2D nanostructures through the effects of sample width, length, substrate interaction and edge disorder.

## Methods

**Sample Fabrication:** Graphene monolayers were deposited on $SiO_2$/Si (~290 nm/0.5 mm) substrates by mechanical exfoliation from natural graphite. Graphene thickness and GNR edge disorder were evaluated with Raman spectroscopy[4, 21, 35]. Samples were annealed in Ar/$H_2$ at 400 °C for 40 minutes. Electron (e)-beam lithography was used to pattern the heater and thermometer electrodes as long, parallel, ~200-nm-wide lines with current and voltage probes, with a separation of $L \approx 260$ nm (Fig. 1). Electrodes were deposited by successive evaporation of $SiO_2$ (20 nm) for electrical insulation and Ti/Au (30/20 nm) for temperature sensing. Additional e-beam lithography and oxygen plasma etching were performed when needed to define GNR arrays with pitch ~150 nm and varying widths.

**Electrical and Thermal Measurements:** The heater electrode is slowly ramped up (<0.2 mHz) to 1.5 mA. We measured the resistance change of the sensor electrode through a lock-in technique with a frequency of 2147 Hz and rms current of 1 μA (carefully verified to avoid additional heating). All electrical measurements were performed in a four-probe configuration, inside a Physical Property Measurement System (PPMS, Quantum Design).

**Numerical Simulation:** We obtain the thermal conductivity by solving the Boltzmann transport equation in the relaxation time approximation including scattering at the rough GNR edges[31].

The simulation uses the phonon dispersion of an isolated graphene sheet, which is a good approximation for $SiO_2$-supported graphene within the phonon frequencies that contribute most to transport[8], and at typical graphene-$SiO_2$ interaction strengths[7]. (However, we note that artificially increasing the graphene-$SiO_2$ coupling, e.g. by applying pressure[43], could lead to modifications of the phonon dispersion and hybridized graphene-$SiO_2$ modes[7].) We assume a graphene monolayer thickness $H$ = 0.335 nm, and a concentration of 1% $^{13}$C isotope point defects[2, 6]. The interaction with the substrate is modeled through perturbations to the scattering Hamiltonian[6] at small patches where the graphene is in contact with the $SiO_2$, with nominal patch radius $a$ = 8.75 nm. Anharmonic three-phonon interactions of both normal and umklapp type are included in the relaxation time (see Supplementary Note 7). An equivalent 2D ballistic scattering rate[25, 29] $\sim 2v_x/L$ is used in the numerical solution ($x$ is direction along graphene) to account for transport in short GNRs.

## Supplementary Information

Supplementary data, error analysis and numerical model details are provided along with the manuscript for review.

## Acknowledgements

We thank D. Estrada, B. Howe and A. Bezryadin for assistance with the experimental setup, A. Serov for the numerical phonon dispersion, and J. Seol. and L. Shi for providing the original data of Ref. [6]. Experiments were carried out in part in the Frederick Seitz Materials Research Laboratory at the University of Illinois. This work was sponsored by a Presidential Early Career (PECASE) award from the Army Research Office (E.P.), the Office of Naval Research (ONR), the Nanotechnology Research Initiative (NRI), the National Science Foundation (NSF), and a NSF CI TraCS Postdoctoral Fellowship (Z.A.).

## Author Contributions

M.H.B. and E.P. conceived the experimental design. M.H.B. fabricated devices and performed measurements with assistance from Z.L. and Z.Y.O.. Z.L. extracted thermal transport data from the measurements and carried out all uncertainty analysis. Z.A. and P.N.M. performed Boltzmann transport simulations with input from Z.L., E.P. and I.K.. F.X. obtained all AFM images.

E.P., Z.L., M.-H.B. and Z.A. co-wrote the manuscript, with input from all authors.

## Author Information

The authors declare no competing financial interests. Correspondence and requests for materials should be addressed to E.P. (epop@illinois.edu).

## Figure Legends

**Figure 1 | Measurement of heat flow in graphene ribbons.** (**a**) Scanning electron microscopy (SEM) image of parallel heater and sensor metal lines with ~260 nm separation, on top of graphene sample (colorized for emphasis). A thin $SiO_2$ layer under the metal lines provides electrical insulation and thermal contact with the graphene beneath (see Methods and Supplementary Note 1). (**b**) Similar sample after graphene etch, serving as control measurement for heat flow through contacts and $SiO_2$/Si underlayers. (**c**) Heater and sensor lines across array of graphene nanoribbons (GNRs). (**d**) Magnified portion of array with GNR widths ~65 nm; inset shows atomic force microscopy (AFM) image of GNRs. Scale bars of (**a-d**) are 2 μm, 1 μm, 2 μm and 1 μm, respectively. (**e**) Three-dimensional (3D) simulation of experimental structure, showing temperature distribution with current applied through heater line. (**f**) Raman spectra of unpatterned graphene sample (bottom curve) and GNRs (upper curves, offset for clarity). Inset shows scaling of Raman D to G peak area vs. GNR width, consistent with the enhanced role of edge disorder in narrower GNRs[14, 21, 35].

**Figure 2 | Thermal conduction scaling in GNRs.** (**a**) Thermal conductance per cross-sectional area ($G/A$) vs. temperature for our GNRs ($L \approx 260$ nm, $W$ as listed, see Fig. 3c), a "short" unpatterned sample ($L \approx 260$ nm, $W \approx 12$ μm, see Fig. 3b), and a "large" sample from Ref. [6] ($L \approx 10$ μm, $W \approx 2.4$ μm, see Fig. 3a). The short but wide graphene sample attains up to ~35% of the theoretical ballistic heat flow limit[22-24] (also see Supplementary Fig. S9). (**b**) Thermal conductivity for the same samples as in (**a**) (also see Fig. S10). (**c**) Thermal conductivity reduction with length for "wide" samples ($W \gg \lambda$), compared to the ballistic limit ($k_{ball} = G_{ball}L/A$) at several temperatures. Symbols are data for our "short" unpatterned graphene samples (Fig. 1a and Fig. 3b), and "large" samples of Ref. [6] (Fig. 3a). Solid lines are model from eq. 1. (**d**) Thermal conductivity reduction with width for GNRs, all with $L \approx 260$ nm (Fig. 1c,d and Fig. 3c). Solid symbols are experimental data from (**b**), open symbols are interpolations for the listed temperature; lines are fitted model from eq. 2. The thermal conductivity of plasma-etched GNRs in this work appears lower than that estimated for GNRs from unzipped nanotubes[13] at a given width, consistent with a stronger effect of edge disorder[35]. Also see Supplementary Fig. S11.

**Figure 3 | Schematic of size effects and different heat flow regimes.** (**a**) Diffusive heat transport in "large" samples with dimensions much greater than the intrinsic phonon mfp ($L$, $W \gg \lambda$). This regime corresponds to the samples measured in both substrate-supported[6] and suspended graphene[2, 3] studies to date. **(b)** Quasi-ballistic heat flow in "short but wide" samples ($L \sim \lambda$ and $W \gg \lambda$). These correspond to our geometry shown in Fig. 1a, with $L \approx 260$ nm and $W \approx 12$ μm. **(c)** Return to a diffusive heat transport regime as the sample width is narrowed down, and phonon scattering with edge roughness (of rms Δ) begins to dominate. These correspond to our arrays of GNRs from Fig. 1c-e ($L \approx 260$ nm and $W$ varying from 45 nm to 130 nm). A fourth regime (long $L$, narrow $W$) is not shown here, but it can be easily understood from the above.

**Figure 4 | Insights from numerical simulations.** (**a**) Comparison of Boltzmann transport model (lines) with experimental data (symbols, same as Fig. 2b, but here on linear scale). (**b**) Computed scaling of phonon mean free path (mfp) vs. width, for two sample lengths as listed. LA and TA phonons with long intrinsic mfp are subject to stronger size effects from edge scattering than ZA modes, which are primarily limited by substrate scattering. (**c**) Estimated contribution of modes to thermal conductivity vs. edge roughness Δ, for a wide sample and a narrow GNR. Like $W$, changes in Δ also more strongly affect LA and TA modes, until substrate scattering begins to dominate. (**d**) Phonon mfp vs. energy (frequency) for a large sample and a narrow GNR. Low-frequency modes are strongly affected by substrate scattering, such that effects of edge roughness are most evident for $\hbar\omega > 15$ meV. Panels (**b-d**) are all at room temperature. Also see Supplementary Figs. S13 and S15.

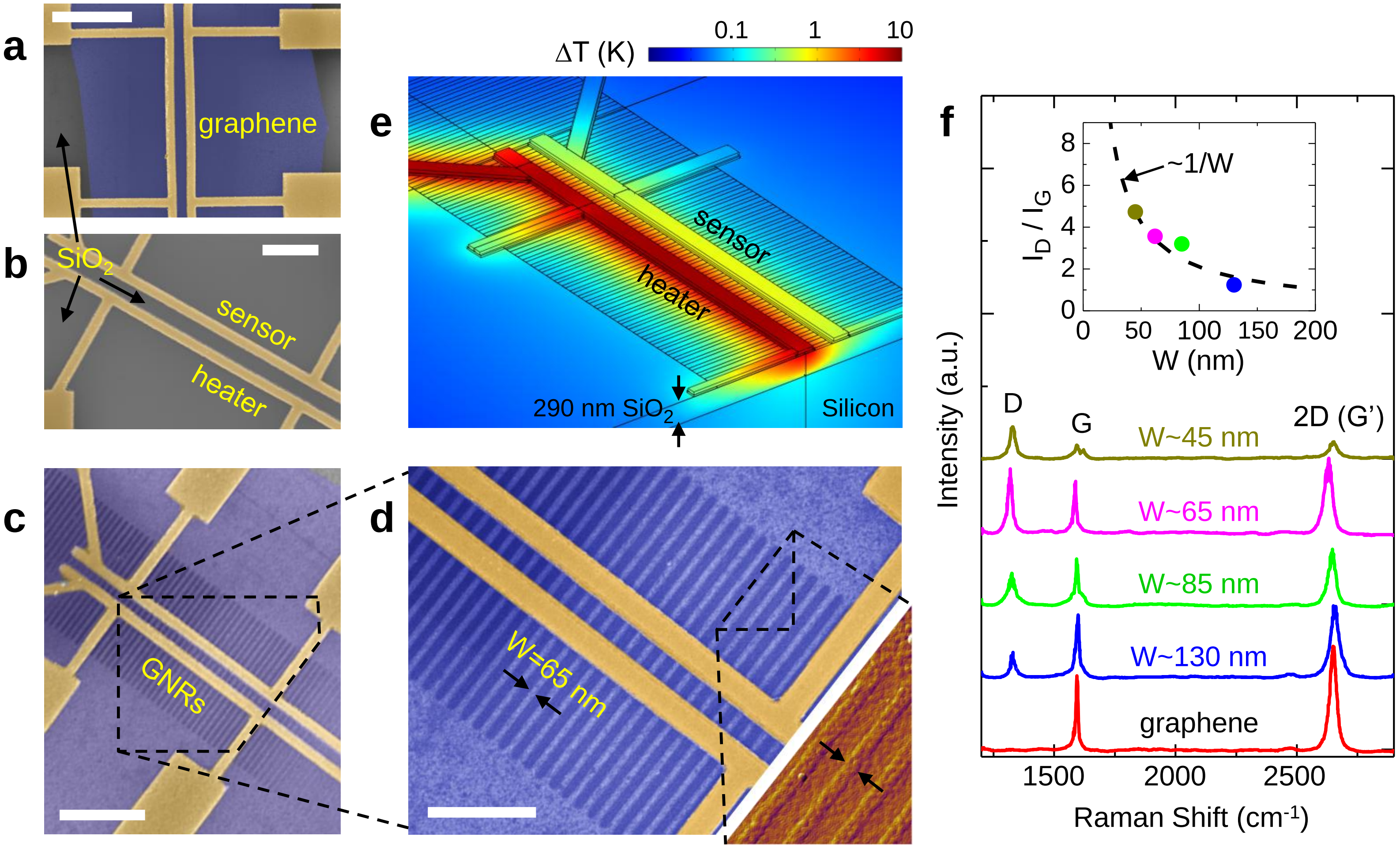

# Figure 1

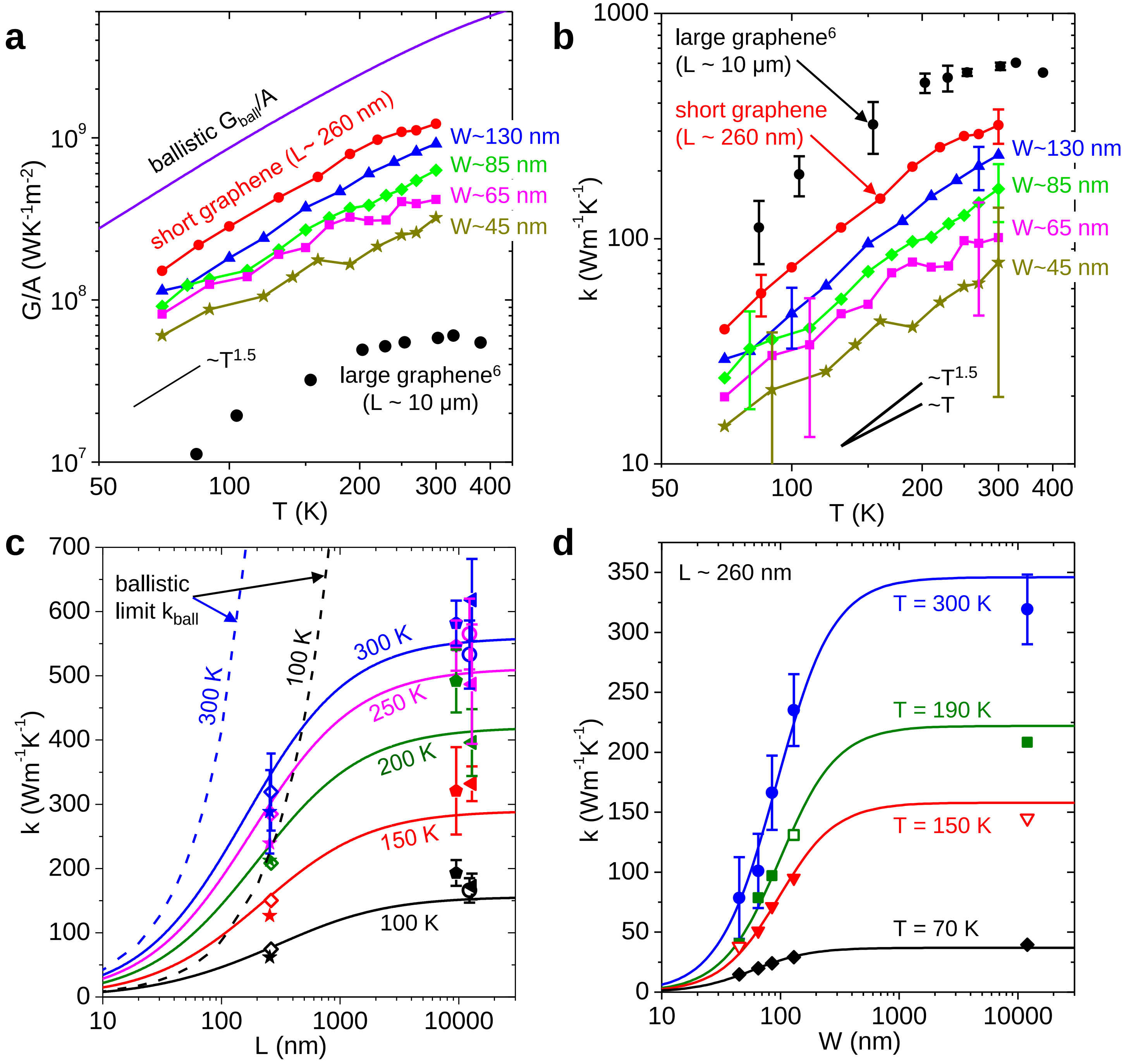

# Figure 2

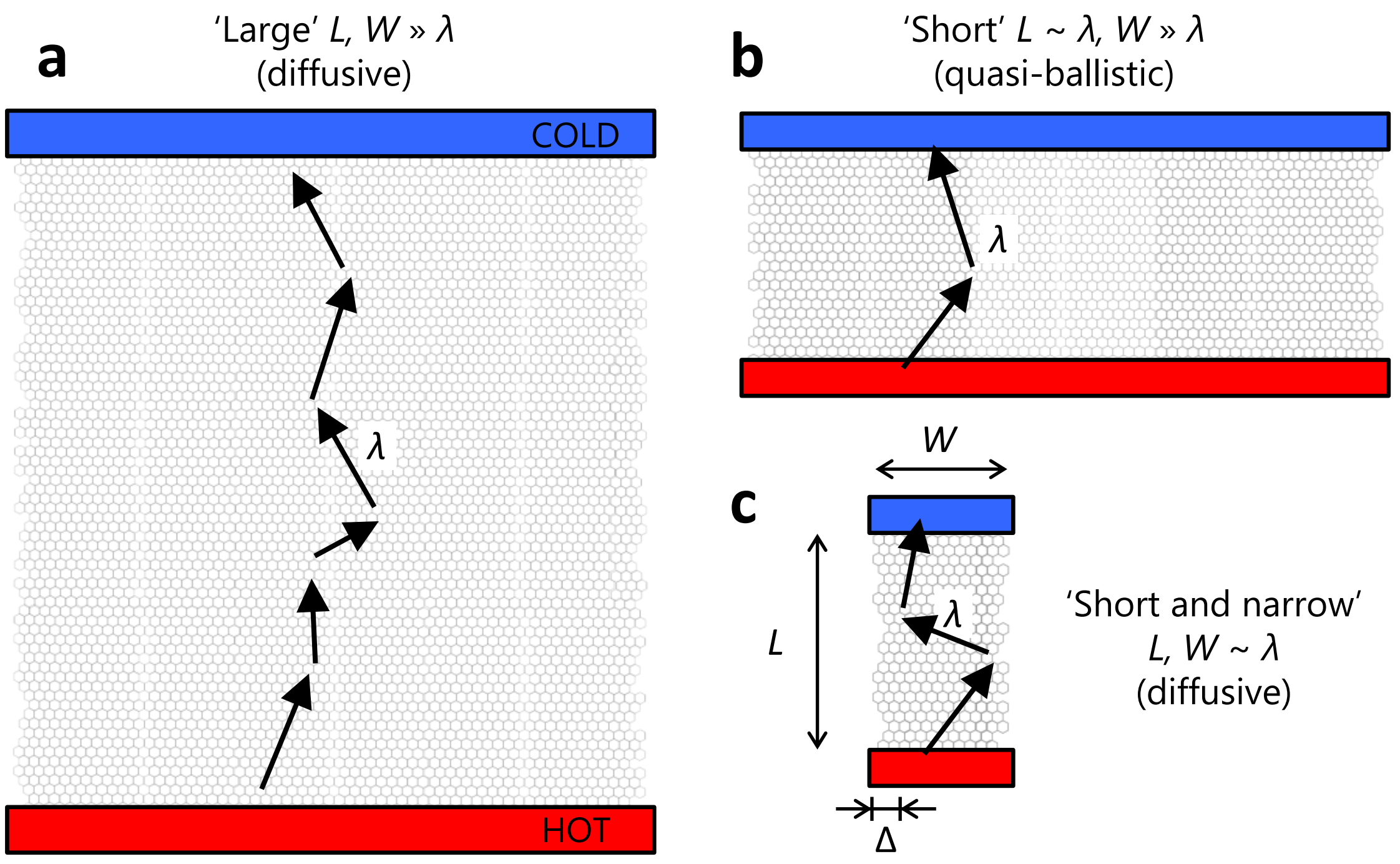

# Figure 3

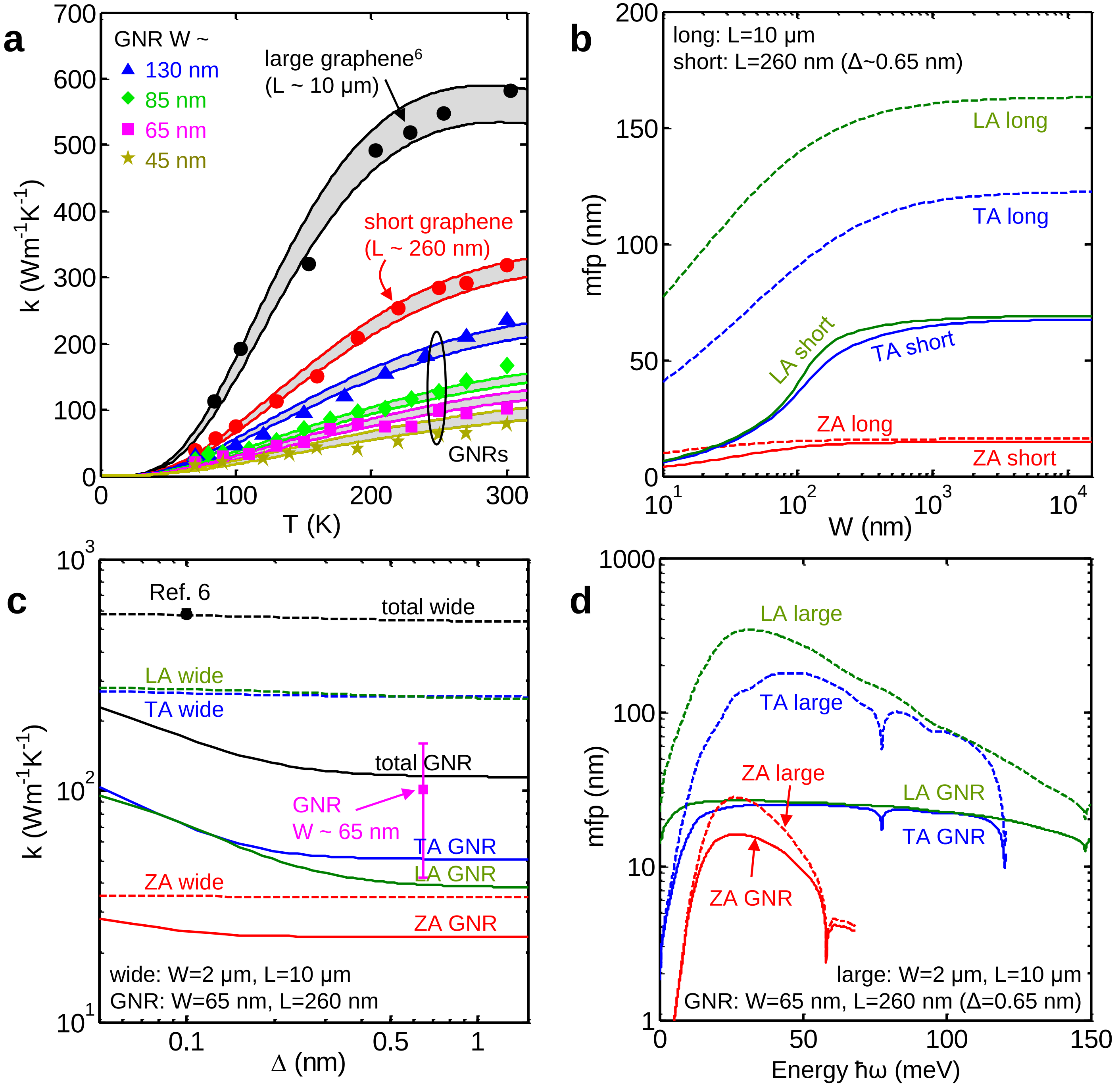

# Figure 4

# Supplementary Information

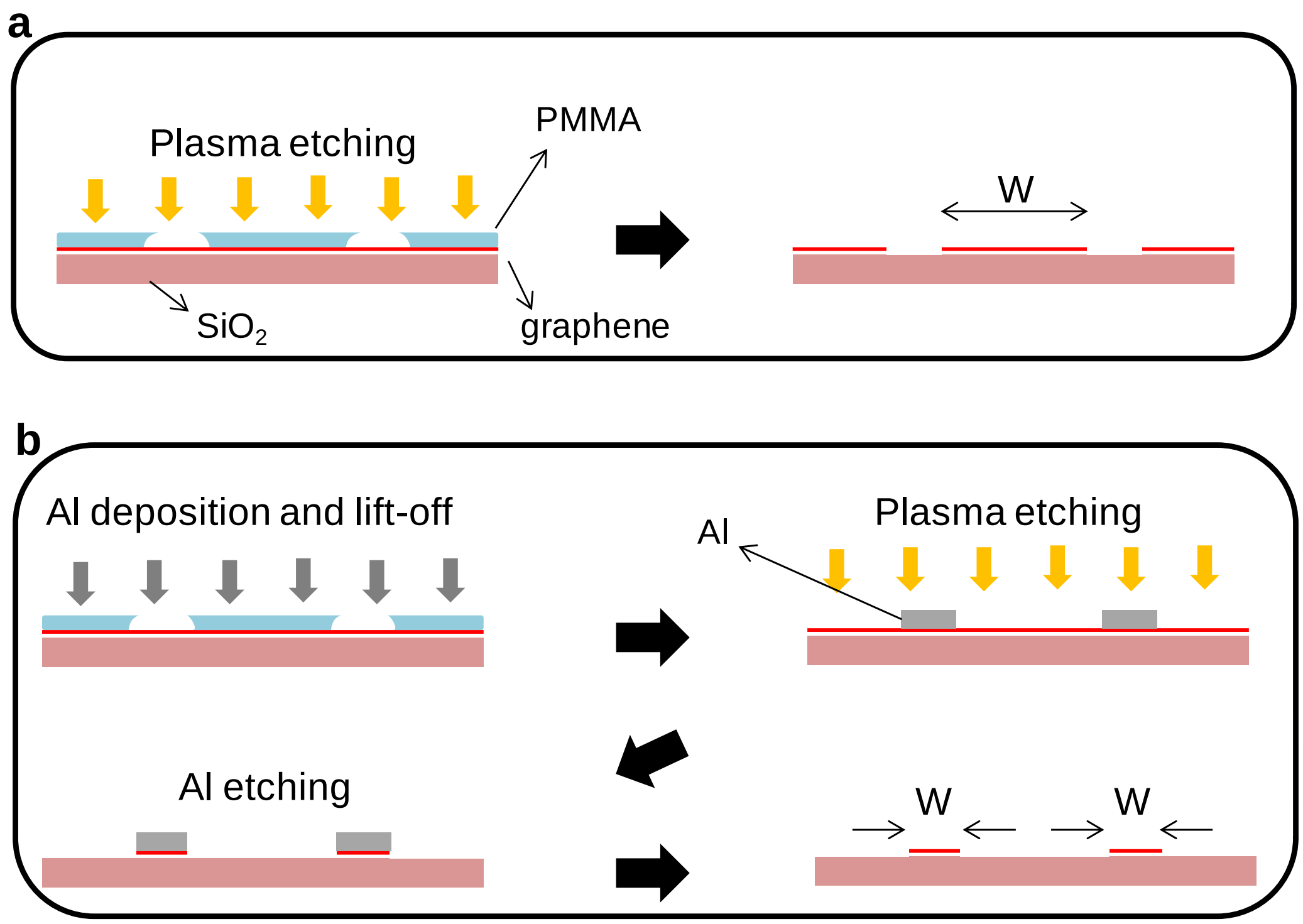

**Supplementary Figure S1 | Process to define the graphene nanoribbon (GNR) widths.** (**a**) PMMA mask method. (**b**) Al mask method.

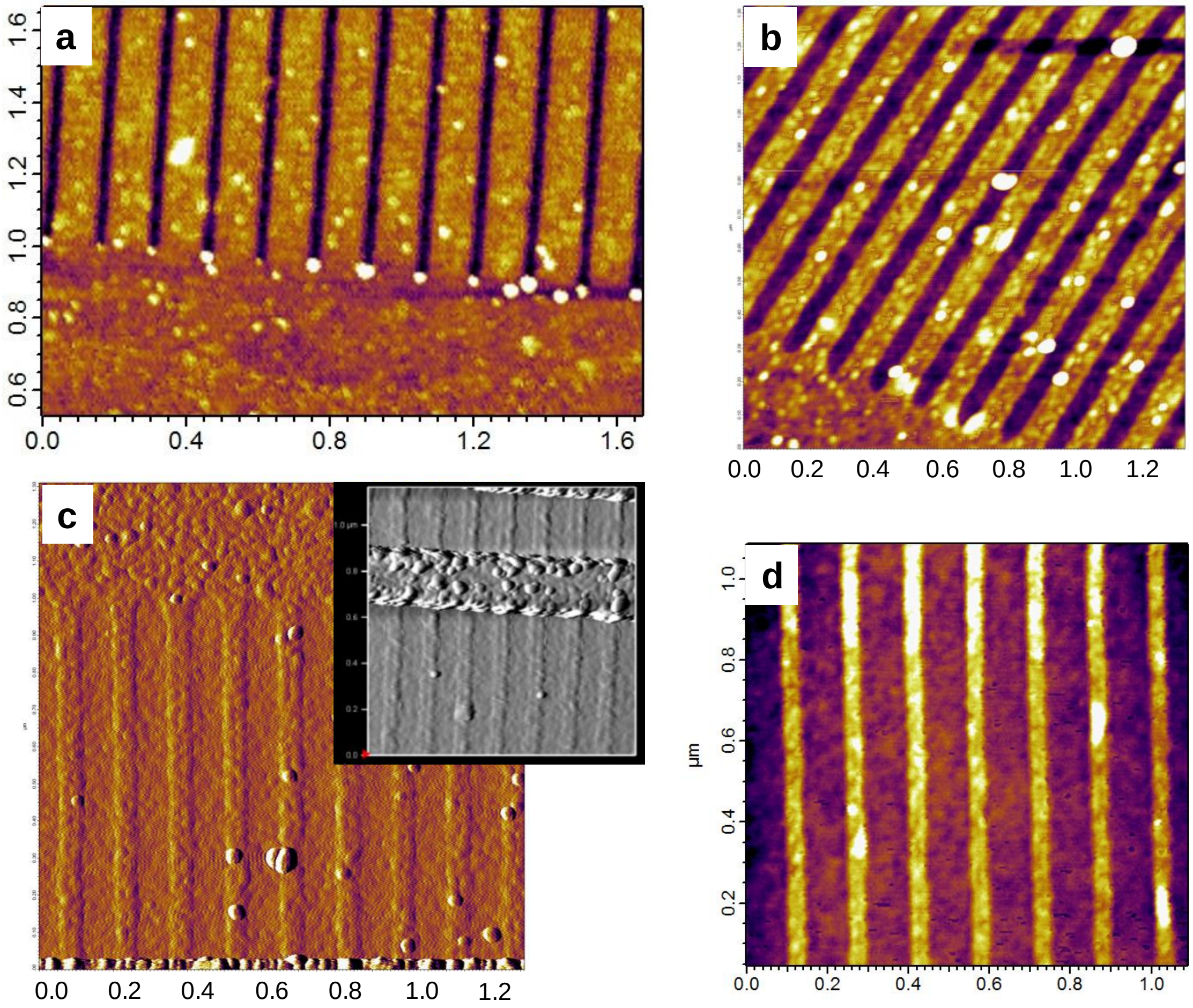

**Supplementary Figure S2 | Atomic force microscopy (AFM) images of GNR arrays.** (**a**) *W* ~ 130 nm GNRs. (**b**) *W* ~ 85 nm GNRs. (**c**) *W* ~ 65 nm GNRs. Inset: AFM image near metal electrodes. (**d**) *W* ~ 45 nm GNRs. The axis units are given in microns on each panel.

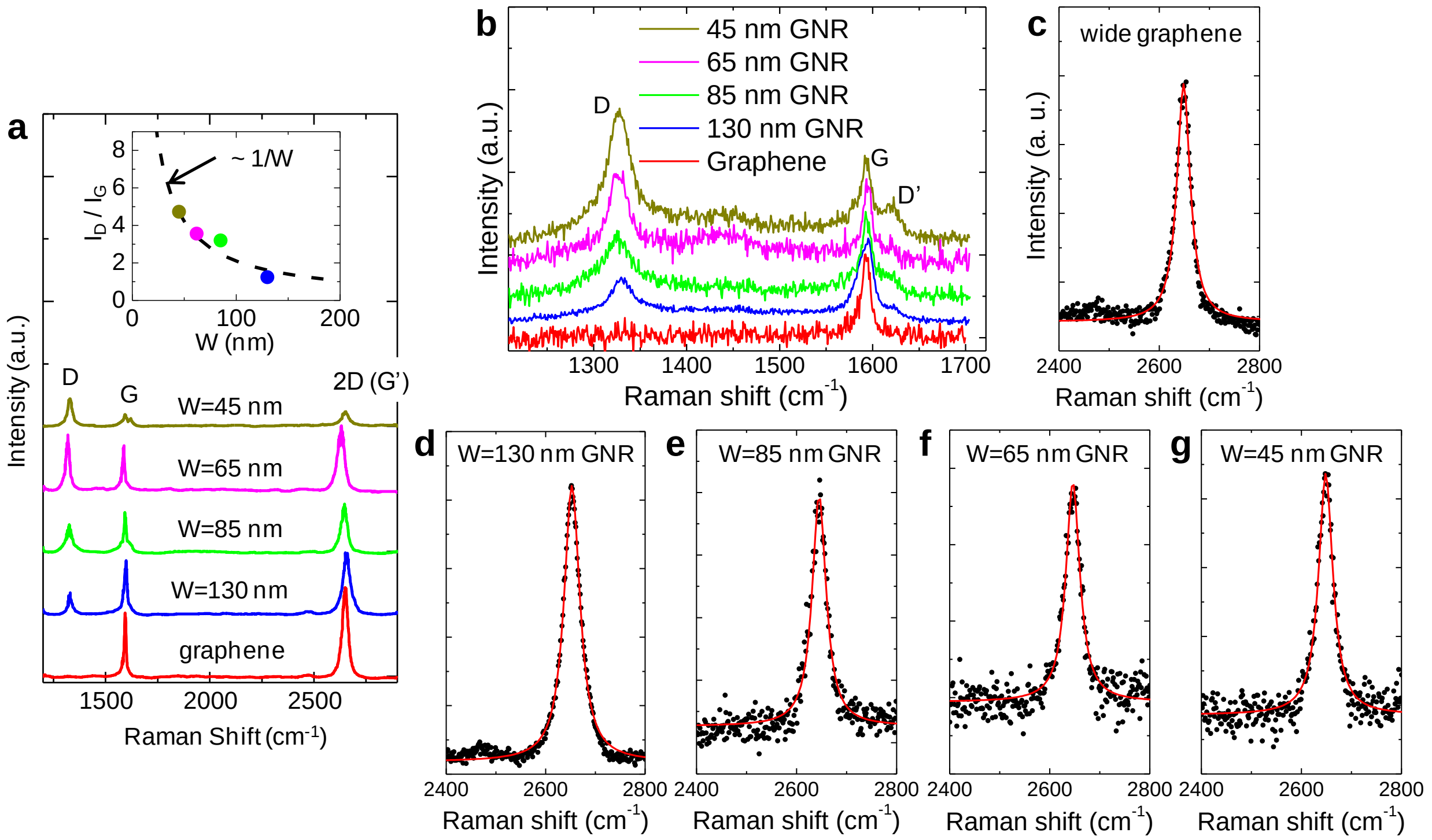

**Supplementary Figure S3 | Raman spectra of GNR arrays and un-patterned graphene.** (**a**) Raman signal for $W$ ~ 130, 85, 65, 45 nm GNRs and un-patterned graphene (same as Fig. 1f of main text, repeated here for convenience). Each spectrum is vertically offset for clarity. Inset is the $I_D/I_G$ ratio as a function of GNR width. (**b**) Zoomed-in D, G, and D' bands of all samples. (**c-g**) 2D bands with a single Lorentzian fit for all samples, consistent with the existence of monolayer graphene.

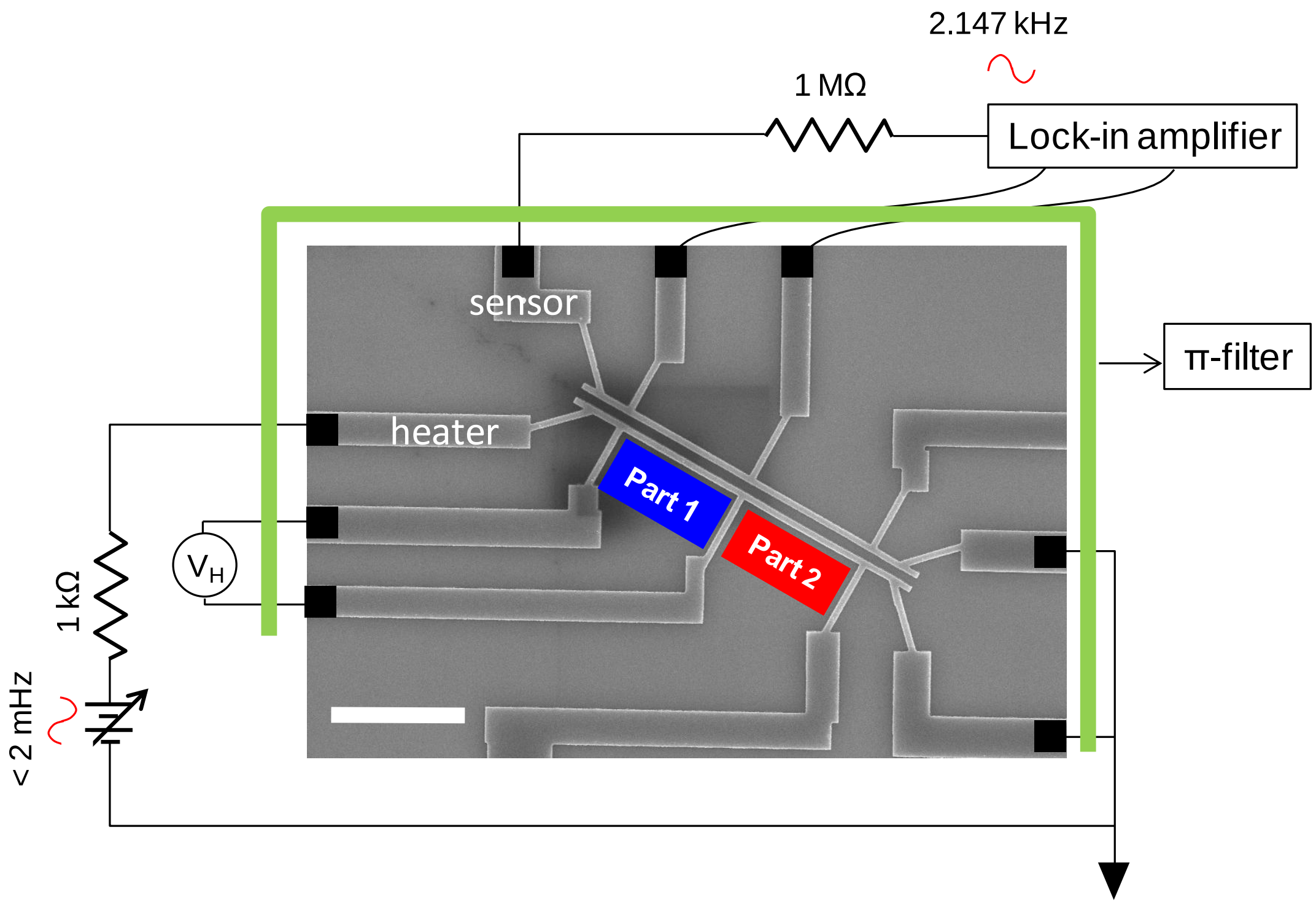

**Supplementary Figure S4 | Scanning electron micrograph (SEM) of thermometry platform and measurement configuration.** Scale bar is 4 μm. Image taken of sample after graphene was etched off, and after all electrical and thermal measurements were completed. Dark region around "part 1" is substrate charging due to previous SEM imaging performed to obtain Fig. 1b in the main text.

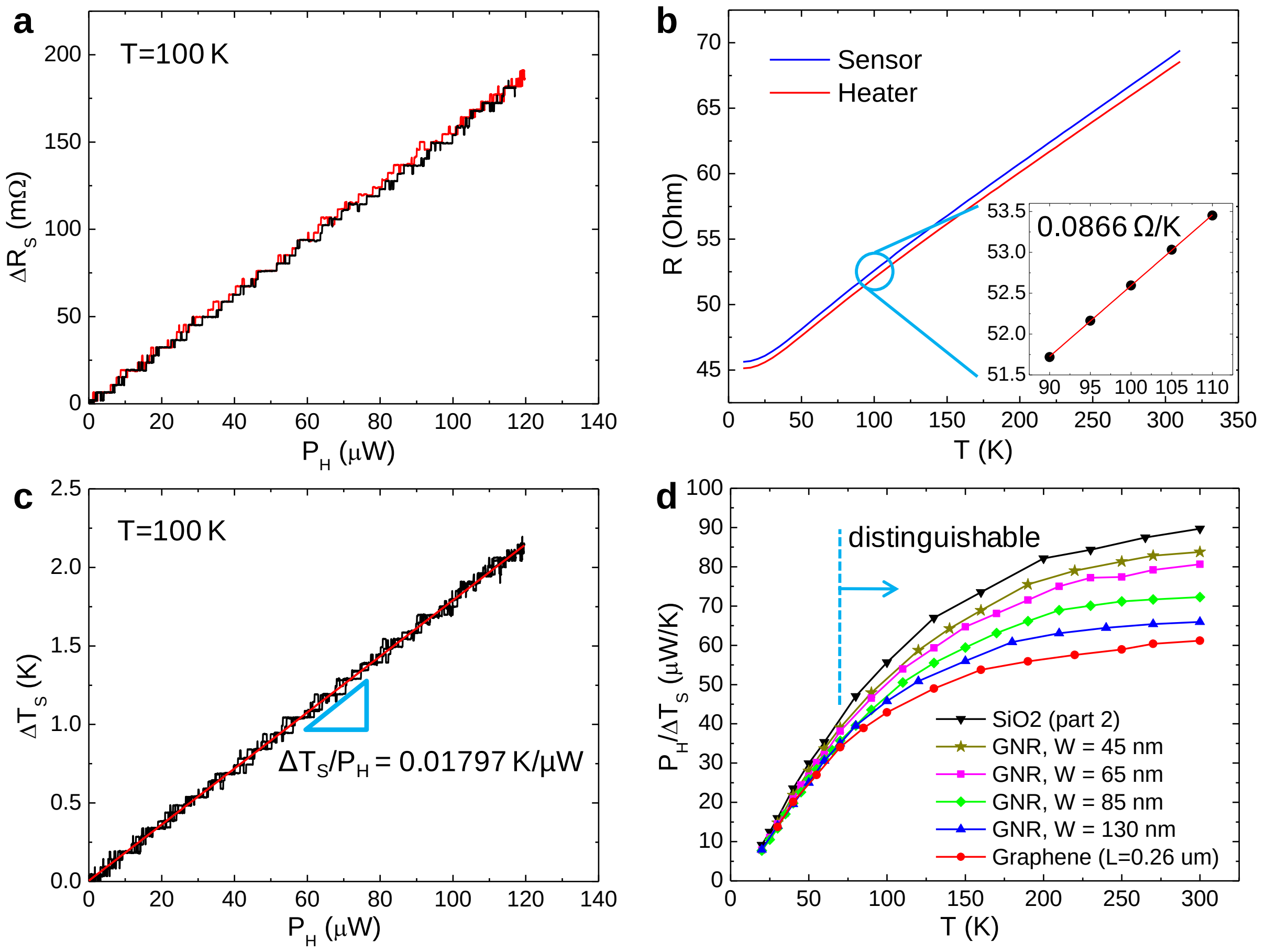

**Supplementary Figure S5 | Measurement process.** (**a**) Sensor resistance change, $\Delta R_S$ as a function of heater power, $P_H$ at $T$ =100 K for the $SiO_2$ sample (Fig. S4). Red and black lines are taken with current flow in opposing direction. (**b**) Calibration of sensor and heater resistances as a function of temperature. The inset shows the $R$-$T$ curve and slope of the sensor near $T$ = 100 K. (**c**) Converted sensor temperature rise, $\Delta T_S$ as a function of heater power, $P_H$ at $T$ = 100 K from (**a**) and (**b**). The slope of the fitted red line is $\Delta T_S/P_H$ = 0.01797 K/μW, which is later used to extract the thermal properties of the $SiO_2$ layer (see Figs. S7-S8). (**d**) Measured ratio of heater power to sensor temperature rise for all representative samples. The uncertainty of these data is ~2% (Tables S1 and S2), comparable to the symbol size. Although this plot shows all raw data taken, the values can be distinguished without ambiguity only at $T \geq 70$ K, which is the temperature range displayed in the main text Figs. 2 and 4.

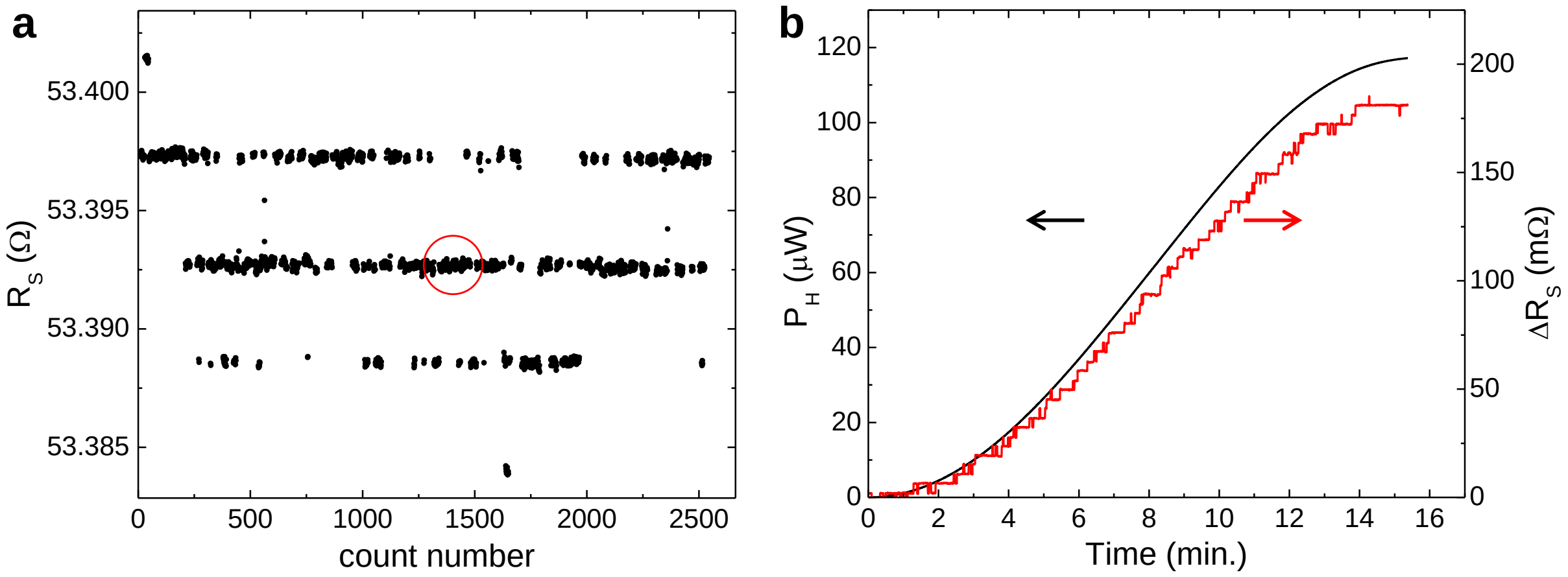

**Supplementary Figure S6 | Measurement error and thermal steady-state.** (**a**) Sensor resistance as a function of count number (time) at background $T$ = 102 K. (**b**) Heater power, $P_H$ and corresponding resistance change in sensor, $\Delta R_S$ as a function of time.

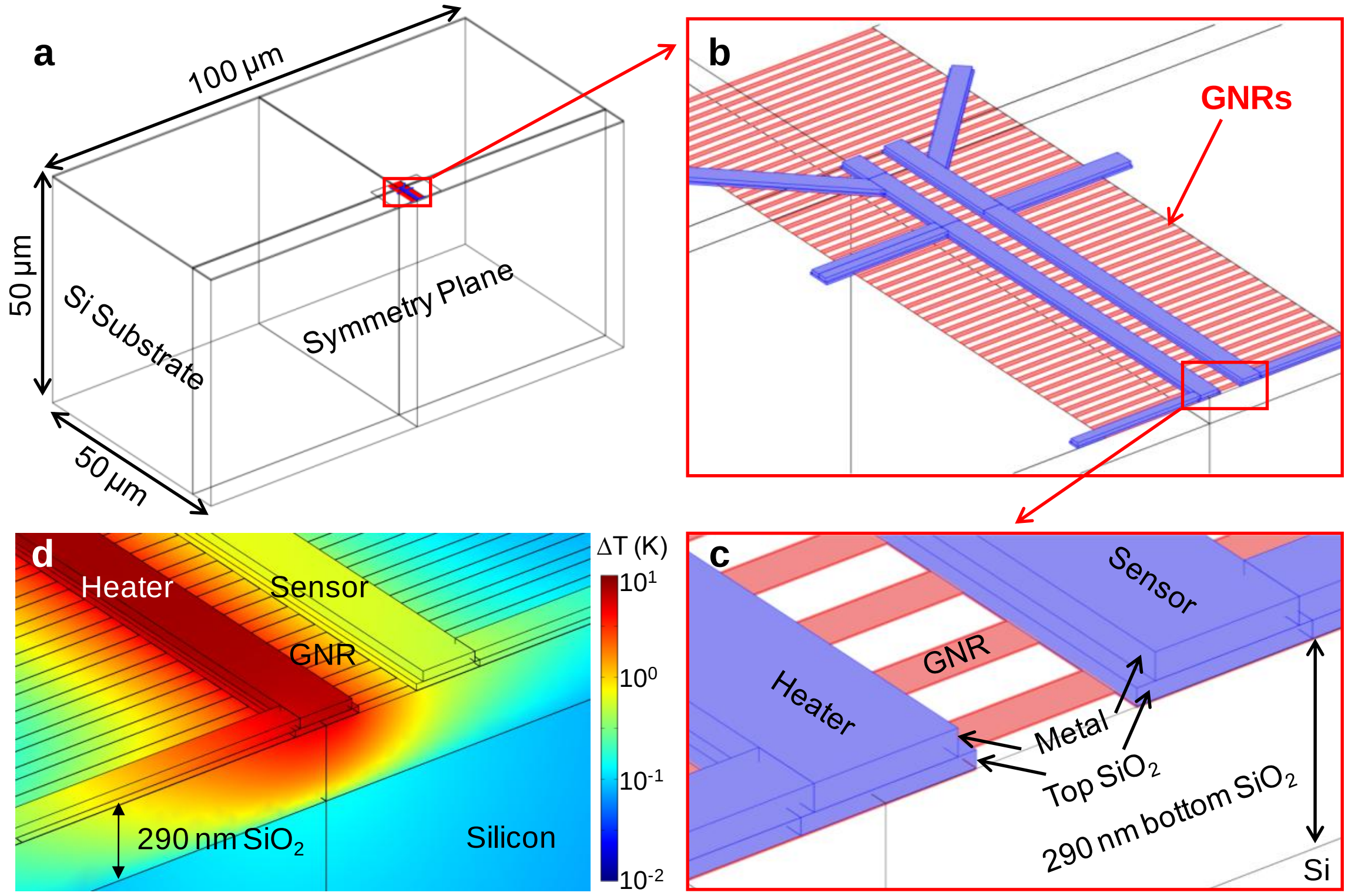

**Supplementary Figure S7 | 3D Finite element method (FEM) model.** (**a**) Whole structure of 3D FEM model. (**b**) Zoomed-in structure to show the core area of the thermometry. (**c**) More zoomed-in structure to show different layers. (**d**) Typical distribution of temperature rise due to heating in simulations which matches with measurements.

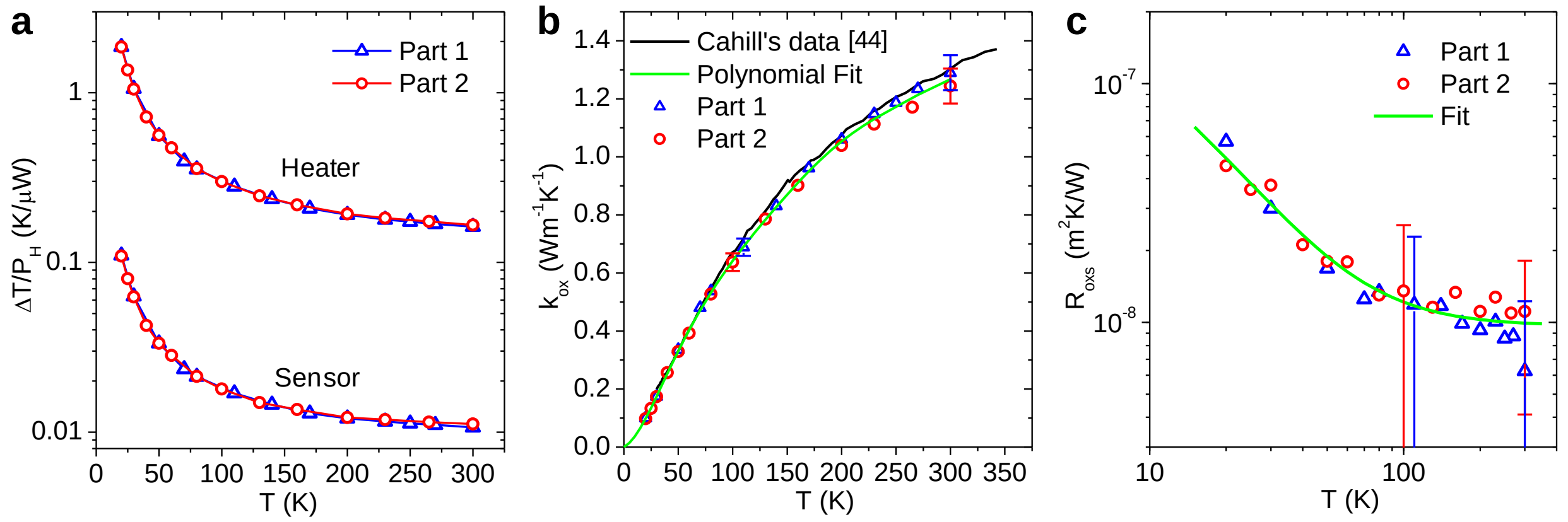

**Supplementary Figure S8 | Control experiment to extract $SiO_2$ thermal properties.** (**a**) Sensor and heater temperature rise normalized by heater power from measurements taken at part 1 and part 2 of the $SiO_2$ sample (see Fig. S4). (**b**) Extracted thermal conductivity of $SiO_2$ from two measurements compared with well-known data from Ref. 44. The green solid line is the polynomial fit up to the 7th order to our data. (**c**) Extracted thermal boundary resistance of the $SiO_2$-Si interface, $R_{oxs}$. The green solid line is the fit to our data by Eq. S1.

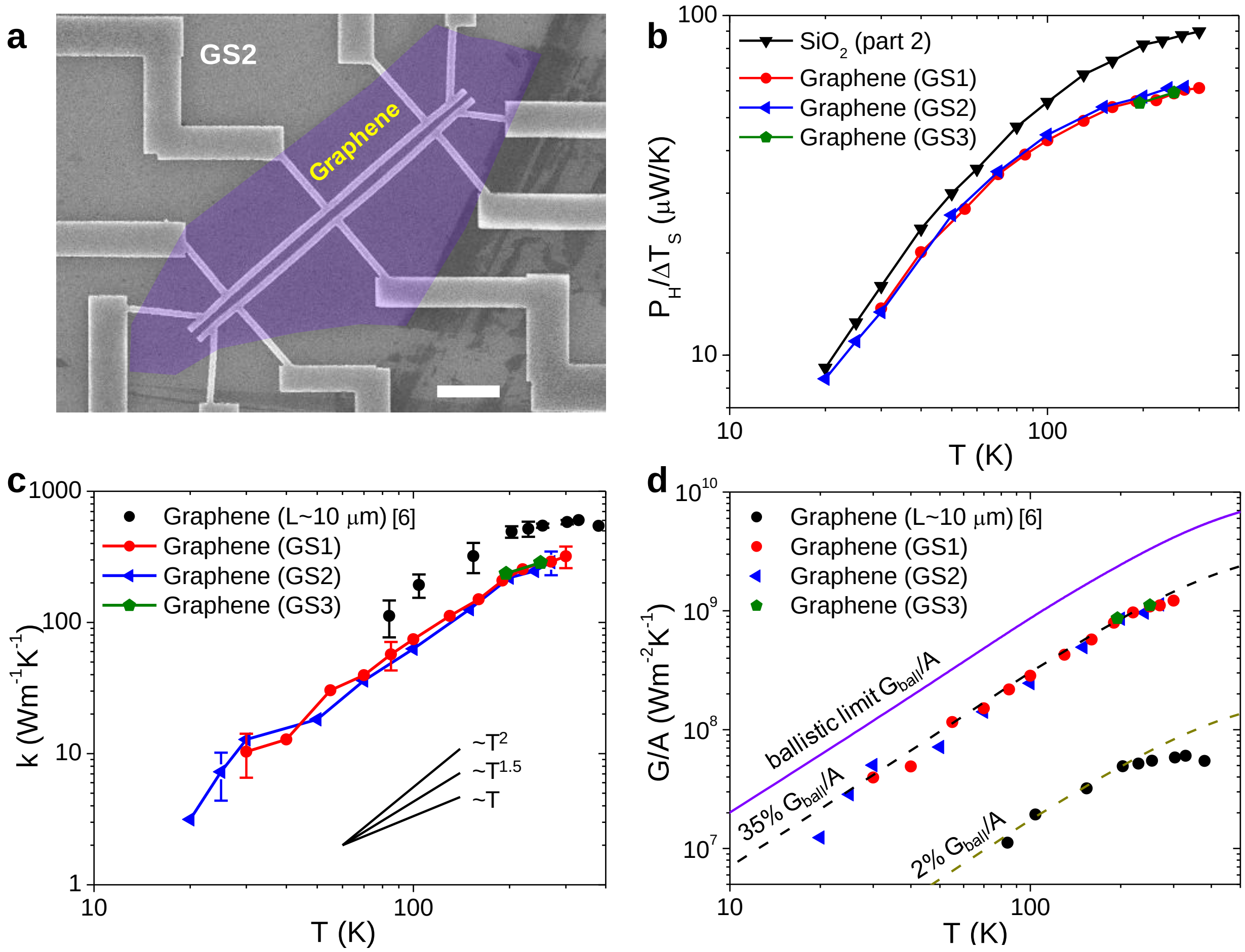

**Supplementary Figure S9 | Data of three unpatterned graphene samples (all *L* ~ 260 nm).** (**a**) SEM image of the unpatterned graphene sample GS2 (colorized for emphasis). (**b**) Measured ratio of heater power to sensor temperature rise of all three graphene samples compared with that of $SiO_2$ sample from the control experiment (Fig. S8). (**c**) Extracted graphene thermal conductivity of all our "short" samples (*L* ~ 260 nm) compared with the "long" graphene (*L* ~ 10 µm) reported in Ref. 6. (**d**) Thermal conductance per unit area compared to theoretical ballistic limit of graphene.

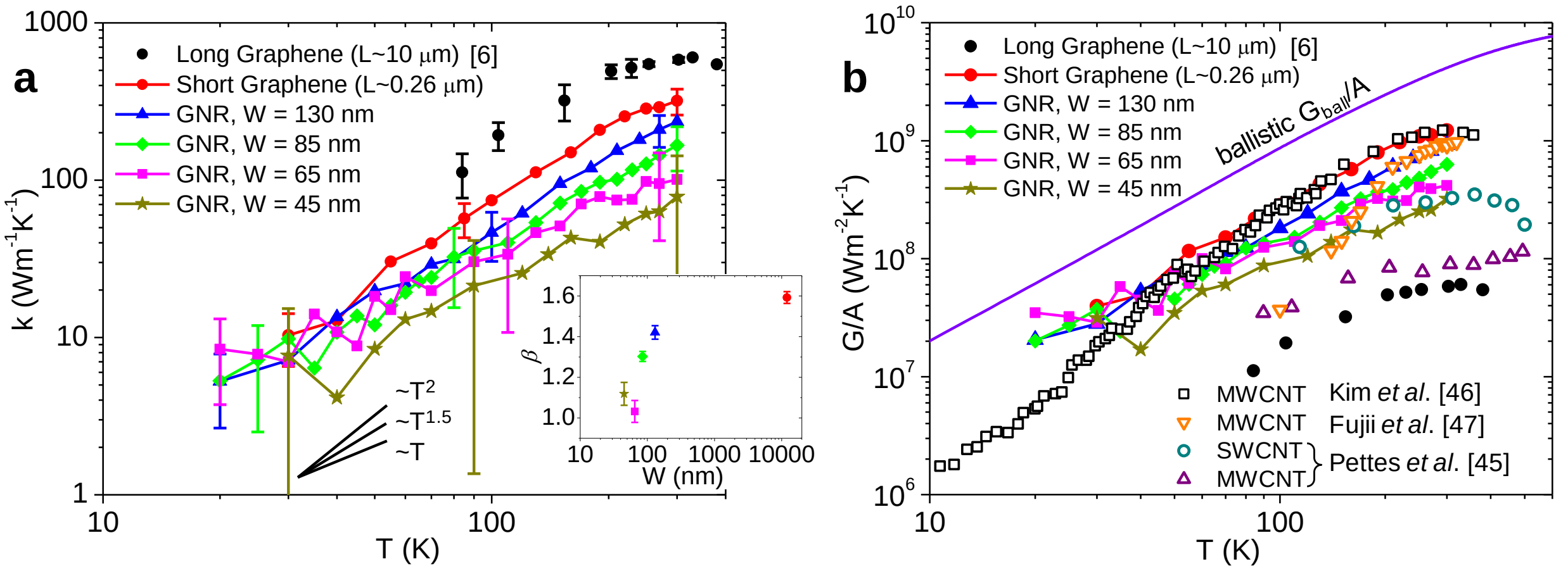

**Supplementary Figure S10 | Complete data sets including low-*T* range and comparison with CNTs.** (**a**) Extracted thermal conductivity of graphene (GS1) and GNRs, same as Fig. 2 in main text but with low-*T* (20~60 K) data included for completeness. Long graphene data are from Ref. 6. Inset: power exponent $\beta$ vs. GNR width fit from $k = \alpha T^{\beta}$ (see text). (**b**) Thermal conductance ($G/A$) of our data compared with those of long graphene[6], ballistic limit, SWCNT of M. Pettes *et al.*[45], and MWCNTs from studies of P. Kim *et al.*[46], M. Pettes *et al.*[45], and M. Fujii *et al.*[47].

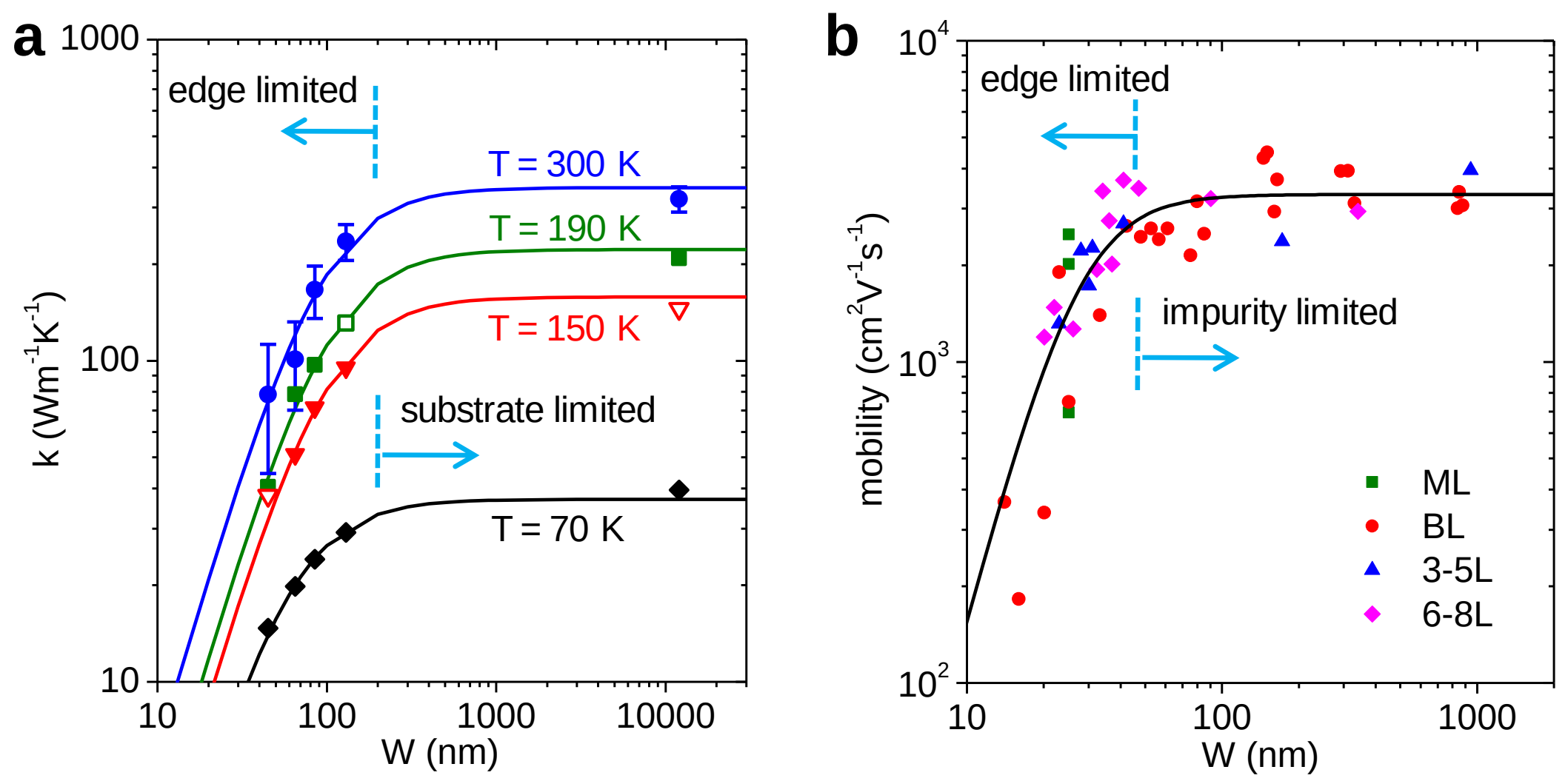

**Supplementary Figure S11 | Width dependence of GNR thermal conductivity and electrical mobility.** (**a**) Thermal conductivity *k* vs. *W* at several temperatures from this study (L ~ 260 nm); same plot as Fig. 2d but on a log-log scale. (**b**) Electrical mobility $\mu$ vs. *W* at room temperature for several layer (L) GNRs, adapted from Ref. 11. Solid line is a fit to data points with $\mu = (1/0.0163W^3 + 1/3320)^{-1}$. Both data sets show a decreasing trend as *W* is narrowed, but with different fall-offs (see text).

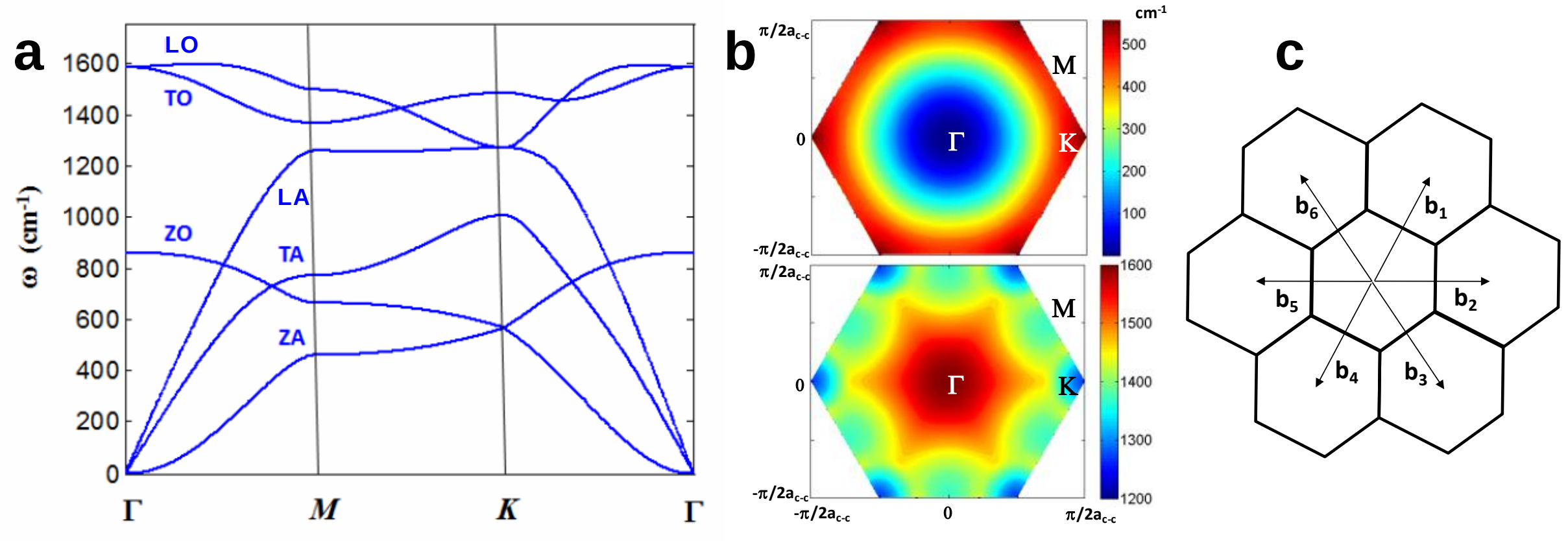

**Supplementary Figure S12 | Phonon dispersion of graphene.** (**a**) Computed dispersion relation in graphene based on force constant model where force constants are approximated up to the 4th nearest neighbors. Six branches are observed: out of plane acoustic (ZA), in plane transverse acoustic (TA), in plane longitudinal acoustic (LA), out of plane optical (ZO), in plane transverse optical (TO) and in plane longitudinal optical (LO). (**b**) Dispersion of ZA mode (top) and TO mode (bottom). (**c**) Representation of graphene reciprocal lattice with basis vectors to the first nearest reciprocal cells.

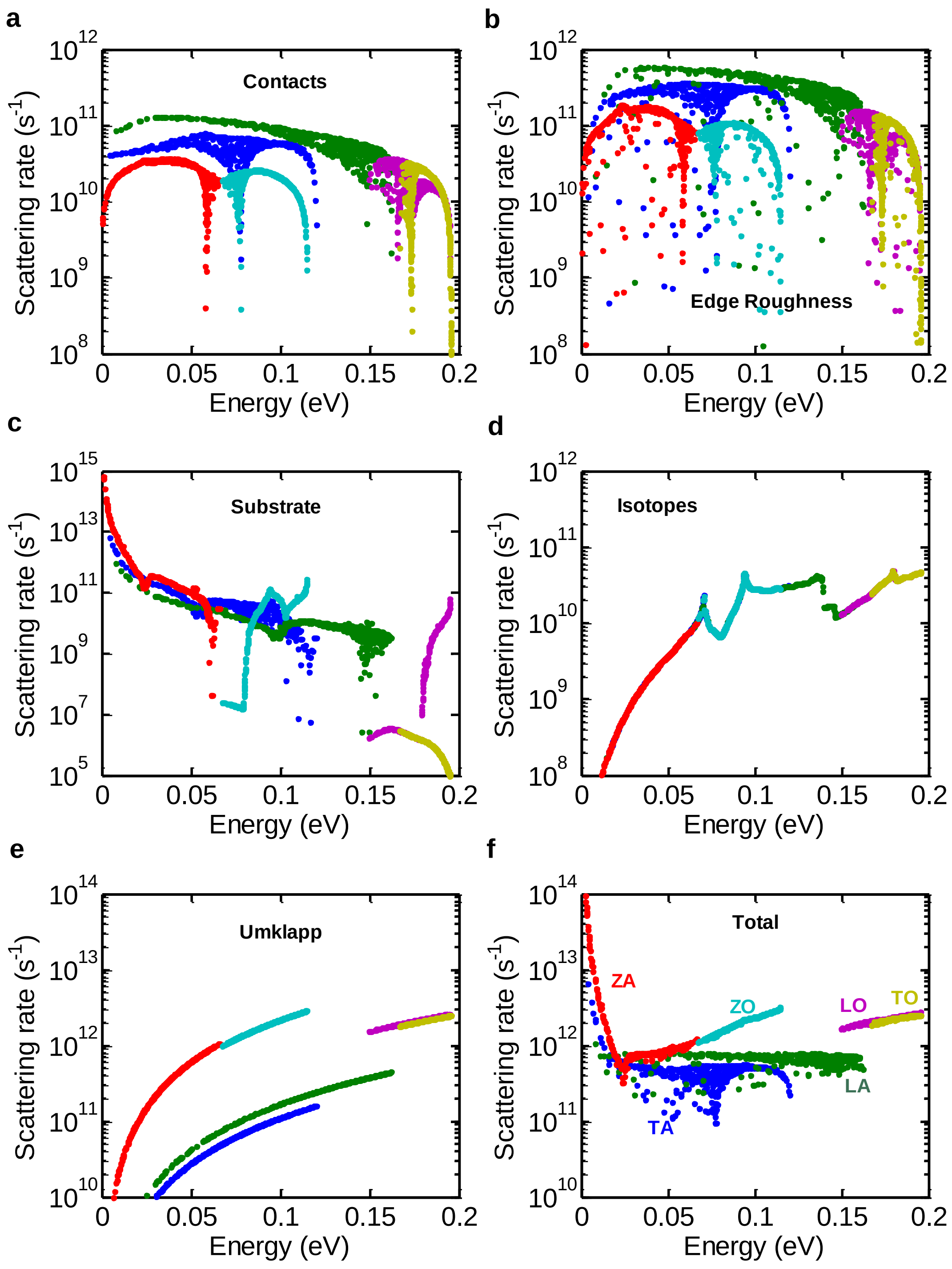

**Supplementary Figure S13 | Computed phonon scattering rates at $T$ = 300 K for a GNR with $W$ = 65 nm, $L$ = 260 nm, and $\Delta$ = 0.65 nm.** Substrate contact patch radius is $a$ = 8.75 nm. Rates are plotted as a function of energy for each scattering mechanism and their total; however, each dot represents one phonon mode. Note that angle-dependent mechanisms like contact and edge roughness scattering have additional dependence on the angle of the phonon velocity vector, which can lead to different rates for the same value of phonon energy.

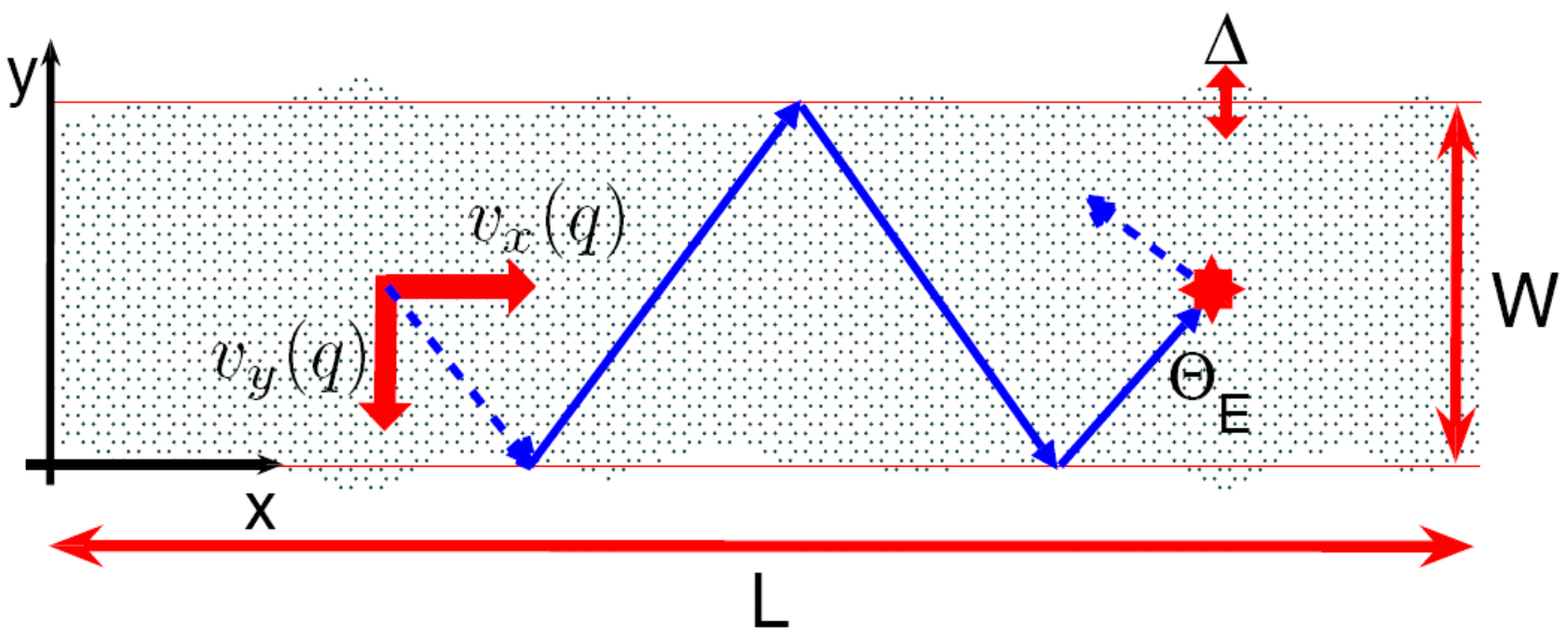

**Supplementary Figure S14 | Schematic of a rough graphene nanoribbon.** The definitions of width (*W*), length (*L*), edge roughness (Δ), and edge angle ($\theta_E$) are indicated. The *x* and *y* coordinates as well as the *x* and *y* components of phonon velocity are also depicted.

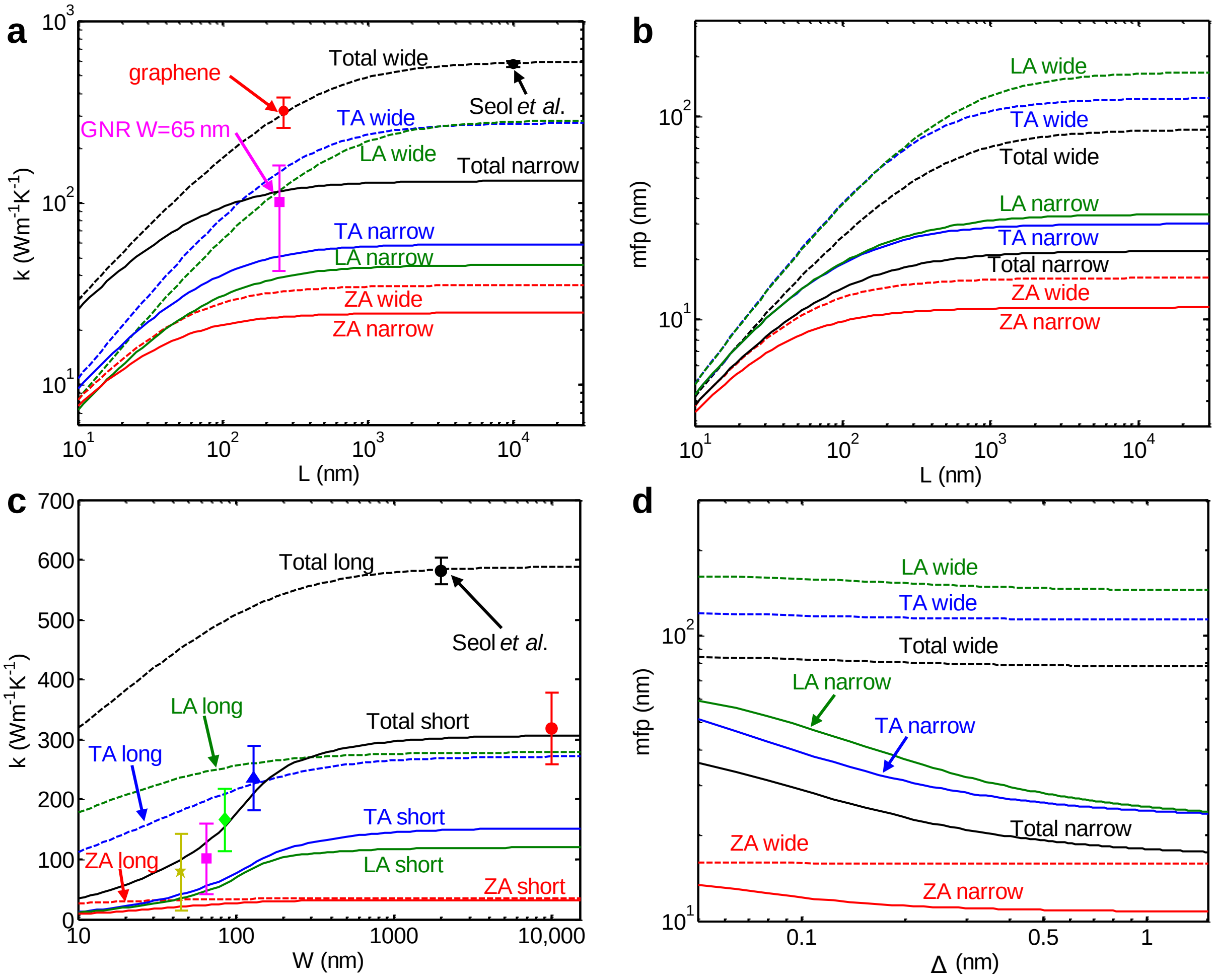

**Supplementary Figure S15 | Dimension scaling behaviors.** (**a**) Variation of lattice thermal conductivity and (**b**) mean free path with the length of graphene samples for a wide (*W*=2 μm) and narrow (*W*=65 nm) ribbon, showing that the length effect is more pronounced in wide ribbons. Phonons in narrow ribbons suffer more scattering at the rough edges; hence, the effect of length is weaker in narrow GNRs. (**c**) Dependence of thermal conductivity on ribbon width shows good agreement with experimental data (symbols). Our model shows that thermal conductivity in short (*L*=260 nm) ribbons is independent of width when ribbons are wide (*W*>*L*), but strongly dependent on width when they are narrow (*W*<*L*), consistent with strong diffuse scattering at the rough edges. (**d**) Effect of rms edge roughness Δ is confirmed in the dependence of the mfp, where we can see that the phonon mfp in wide ribbons is largely independent of Δ, while the mfp in narrow ribbons decreases with increasing Δ, indicating the strong role of edge roughness in thermal transport in narrow GNRs.

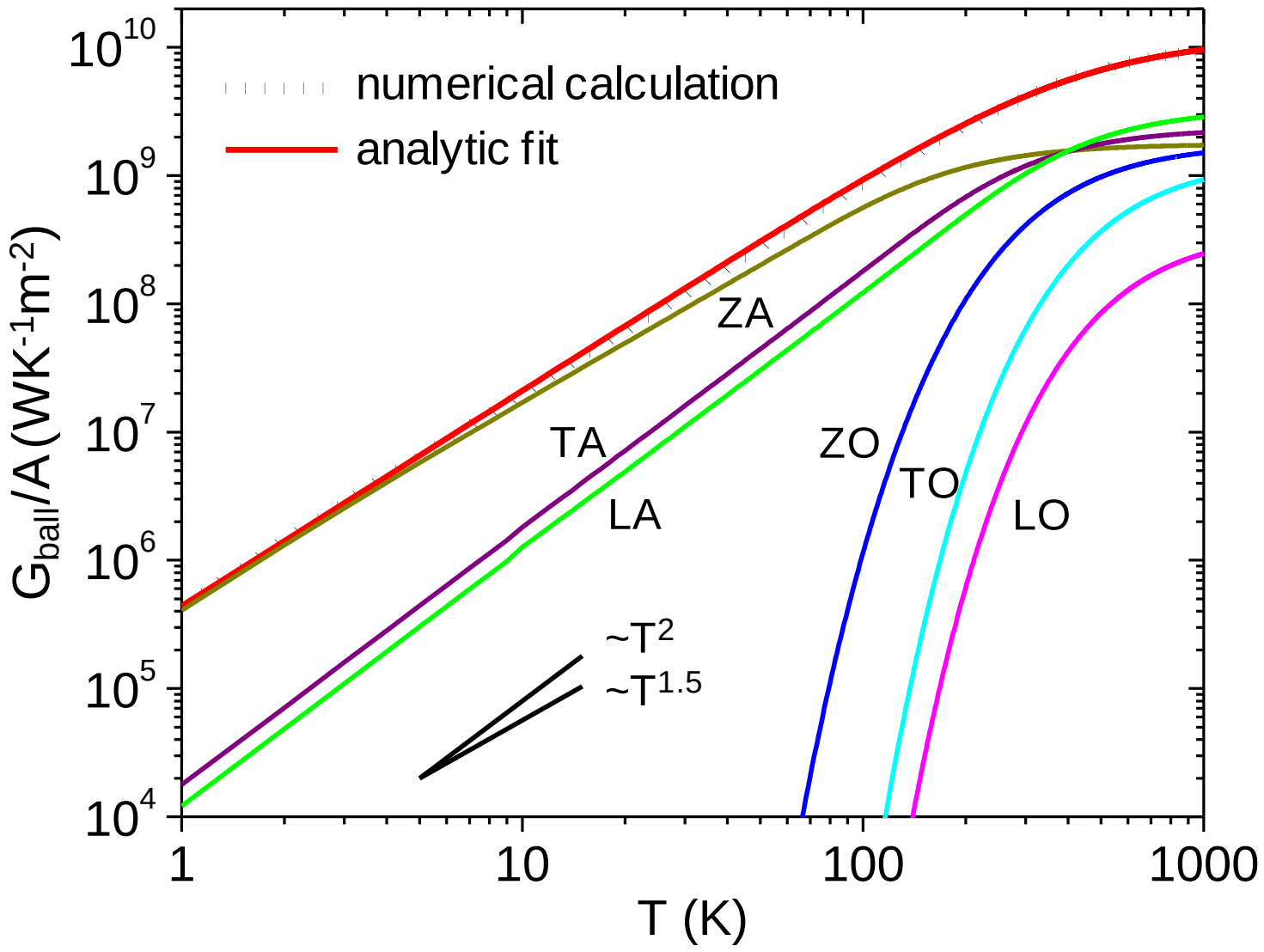

**Supplementary Figure S16 | Ballistic limit of graphene thermal conductance.** Comparison of numerically calculated ballistic conductance $G_{ball}/A$ (symbols) using the full 2D phonon dispersion from Ref. 48, and that of the simple analytic approximation given in Eq. S21 (lines). The ballistic contributions of the various phonon branches are also shown.

**Supplementary Table S1 | Uncertainty analysis for heat flow in the unpatterned "short" graphene.** Example of calculated sensitivities and uncertainty analysis for the thermal conductivity of the graphene sample GS1 at 300 K. The extracted k is 319 $Wm^{-1}K^{-1}$ and its overall uncertainty is 19%.

| Input parameters (T = 300 K) | | | Units | Values $x_i$ | Uncertainty $u_{xi}$ | $u_{xi}/x_i$ | Sensitivity $s_i$ | Contribution $c_i = \lvert s_i \rvert \times u_{xi}/x_i$ | $c_i^2/\Sigma c_i^2$ |
|---|---|---|---|---|---|---|---|---|---|
| Expt. | Sensor response | $\Delta T_S/P_H$ | K/μW | 0.01635 | 0.0002 | 1.2% | 4.49 | 5.5% | 8.6% |
| Thermal | Thermal conductivity of $SiO_2$, Si, metal | $k_{ox}$ | W/m/K | 1.267 | 0.04 | 3.2% | 4.25 | 13.4% | 51.6% |
| | | $k_{Si}$ | | 115 | 10 | 8.7% | 0.55 | 4.8% | 6.6% |
| | | $k_{met}$ | | 55 | 4 | 7.3% | 0.34 | 2.5% | 1.8% |
| | TBR of graphene/$SiO_2$, $SiO_2$/Si, metal/$SiO_2$ interfaces | $R_{gox}$ | $m^2K/W$ | 1.15E-08 | 2.0E-09 | 17.4% | -0.12 | 2.1% | 1.3% |
| | | $R_{oxs}$ | | 9.92E-09 | 3.0E-09 | 30.2% | -0.25 | 7.6% | 16.5% |
| | | $R_{mox}$ | | 1.02E-08 | 3.0E-09 | 29.5% | 0.05 | 1.5% | 0.7% |
| Geometrical | Thickness of bottom and top $SiO_2$, and metal | $t_{box}$ | nm | 288 | 1 | 0.3% | -5.65 | 2.0% | 1.1% |
| | | $t_{tox}$ | | 20 | 1 | 5.0% | -0.01 | 0.1% | 0.0% |
| | | $t_{met}$ | | 50 | 2 | 4.0% | 0.35 | 1.4% | 0.5% |
| | Distance of H/S metal lines and H/S $SiO_2$ lines | $D_{met}$ | | 494 | 5 | 1.0% | 5.38 | 5.4% | 8.5% |
| | | $D_{tox}$ | | 486 | 5 | 1.0% | 2.39 | 2.5% | 1.7% |
| | Width of metal and $SiO_2$ lines | $W_{met}$ | | 186 | 4 | 2.2% | -0.16 | 0.3% | 0.0% |
| | | $W_{tox}$ | | 224 | 4 | 1.8% | -0.17 | 0.3% | 0.0% |
| | Half length of H/S electrodes | $L_{HS}/2$ | μm | 5.86 | 0.03 | 0.5% | 0.04 | 0.0% | 0.0% |
| | Distance of 2 Voltage probes | $D_{pVV}$ | | 4.23 | 0.02 | 0.5% | 3.75 | 1.8% | 0.9% |
| | Distance of Current and Voltage probes | $D_{pIV}$ | | 1.04 | 0.02 | 1.9% | 0.003 | 0.0% | 0.0% |

**Supplementary Table S2 | Uncertainty analysis for GNR thermal conductivity.** Example of calculated sensitivities and uncertainty analysis for the thermal conductivity of GNRs with $W$ = 65 nm at 300 K. The extracted k is 101 $Wm^{-1}K^{-1}$ and its overall uncertainty is 60%.

| Input parameters (T = 300 K) | | | Units | Values $x_i$ | Uncertainty $u_{xi}$ | $u_{xi}/x_i$ | Sensitivity $s_i$ | Contribution $c_i = \|s_i\| \times u_{xi}/x_i$ | $c_i^2/\Sigma c_i^2$ |
|---|---|---|---|---|---|---|---|---|---|
| Expt. | Sensor response | $\Delta T_S/P_H$ | K/μW | 0.0124 | 0.0002 | 1.6% | 13.22 | 21.3% | 12.9% |
| Thermal | Thermal conductivity of $SiO_2$, Si, metal, and outter graphene | $k_{ox}$ | W/m/K | 1.267 | 0.04 | 3.2% | 10.38 | 32.8% | 30.5% |
| | | $k_{Si}$ | | 115 | 10 | 8.7% | 2.61 | 22.7% | 14.6% |
| | | $k_{met}$ | | 22 | 2 | 9.1% | 0.68 | 6.2% | 1.1% |
| | | $k_g$ | | 320 | 60 | 18.8% | 0.001 | 0.0% | 0.0% |
| | TBR of graphene/$SiO_2$, $SiO_2$/Si, metal/$SiO_2$ interfaces | $R_{gox}$ | $m^2K/W$ | 1.15E-08 | 2.0E-09 | 17.4% | -0.04 | 0.6% | 0.0% |
| | | $R_{oxs}$ | | 9.92E-09 | 3.0E-09 | 30.2% | -1.00 | 30.3% | 26.1% |
| | | $R_{mox}$ | | 1.02E-08 | 3.0E-09 | 29.5% | 0.11 | 3.4% | 0.3% |
| Geometrical | Thickness of bottom and top $SiO_2$, and metal | $t_{box}$ | nm | 291 | 1 | 0.3% | -22.51 | 7.7% | 1.7% |
| | | $t_{tox}$ | | 25 | 1 | 4.0% | -0.17 | 0.7% | 0.0% |
| | | $t_{met}$ | | 40 | 2 | 5.0% | 0.70 | 3.5% | 0.3% |
| | Distance of H/S metal lines and H/S $SiO_2$ lines | $D_{met}$ | | 517 | 5 | 1.0% | 14.73 | 14.2% | 5.8% |
| | | $D_{tox}$ | | 509 | 5 | 1.0% | 14.25 | 14.0% | 5.6% |
| | Width of metal lines, top $SiO_2$ lines, and GNRs | $W_{met}$ | | 218 | 4 | 1.8% | -0.64 | 1.2% | 0.0% |
| | | $W_{tox}$ | | 266 | 4 | 1.5% | -2.27 | 3.4% | 0.3% |
| | | $W_{GNR}$ | | 65 | 3 | 4.6% | -0.80 | 3.7% | 0.4% |
| | Half length of H/S electrodes | $L_{HS}/2$ | μm | 5.77 | 0.03 | 0.5% | 0.20 | 0.1% | 0.0% |
| | Distance of 2 Voltage probes | $D_{pVV}$ | | 4.17 | 0.02 | 0.5% | 7.80 | 3.7% | 0.4% |
| | Distance of Current and Voltage probes | $D_{pIV}$ | | 1.04 | 0.02 | 1.9% | -0.286 | 0.5% | 0.0% |

## Supplementary Note 1: Fabrication and Characterization of Graphene Nanoribbons

***Fabrication process:*** We used two approaches to define and fabricate graphene nanoribbons (GNRs): one with a PMMA mask (Fig. S1a), the other with an Al mask (Fig. S1b)[49]. Double poly(methyl methacrylate) (PMMA) layers (PMMA 495K A2/PMMA 950K A4) were coated on the Si/$SiO_2$ substrate. For the electron (e)-beam lithography, we used 30 keV e-beam accelerating voltage. After opening 40 nm wide PMMA windows, we etched the graphene exposed through the windows with an oxygen plasma, creating GNRs of width $W$ (Fig. S1a). This PMMA mask method was used for the $W \approx 130$ nm, ~85 nm, and ~65 nm wide GNRs. For the narrower ~45 nm GNRs we used Al masks (Fig. S1b). In this case, after opening the PMMA windows, instead of plasma etching, we deposited 30 nm thick Al and obtained ~45 nm wide Al strips on graphene. After plasma etching of exposed graphene and Al etching (type A, Transene Company) we obtained ~45 nm wide GNRs. Figure S2 shows the atomic force microscopy (AFM) images of fabricated GNR arrays with $W \approx$ 130 nm, 85 nm, 65 nm, and 45 nm, respectively. The bottom and top regions of Figs. S2a and S2c correspond to the un-etched pristine graphene.

***Raman characterization of GNR arrays*****:** To characterize the prepared GNRs, we performed Raman spectroscopy with a 633 nm wavelength laser (~1 µm spot size) as shown in Fig. S3 and Fig. 1f of the main text. Even before patterning into GNRs, we selected only monolayer graphene flakes, identifiable through their 2D (G') to G Raman peak ratio, and through a single fitted Lorentzian to their 2D (G') peak. The unpatterned graphene samples had no identifiable D peak, indicative of little or no disorder[50]. On the other hand, the GNR arrays showed a pronounced D band consistent with the presence of edge disorder[21]. The peak intensity of the D band with respect to that of the G band increases with narrower GNR width. Because the edges of graphene serve as defects by breaking the translational symmetry of the lattice, the larger fraction of the edge in narrow GNRs will enhance the D peak[51]. The inset of Fig. S3a (same as Fig. 1f in the main text, repeated here for convenience) quantitatively shows the behavior by calculating the ratio of integrated D band ($I_D$) to G band ($I_G$), $I_D/I_G$, as a function of GNR width (symbols). The width dependence of the peak ratio follows a relation of $I_D/I_G = cW^{-1}$ with $c = 210$ nm (dashed line), which is consistent with previous reports of GNR characterization[21, 52]. Figure S3b shows the D, G, and D' peaks in detail of fabricated GNR arrays and un-patterned graphene with 633 nm wavelength laser. Figures S3c-g show the Raman 2D band spectra (scattered points) for the samples. All 2D bands are fit by a single Lorentzian peak (solid red curves) with ~2650 $cm^{-1}$ peak position, which is consistent with previous reports of monolayer graphene and GNRs[21].

## Supplementary Note 2: Experimental Set-Up and Data Analysis

***Experimental set-up*:** Figure S4 shows a scanning electron microscopy (SEM) image of our typical thermometry platform of a substrate-supported sample with heater and thermometer (sensor) electrodes. Here, the sample is shown for the measurement of the $SiO_2$ layer, after the exposed graphene has been removed by an oxygen plasma etch (however, graphene still exists under the metal electrodes, consistent with the other samples). To block environmental noise including electrostatic discharge, π-filters with a cut-off frequency of 2 MHz were inserted across all measurement lines. To control the temperature, PPMS (Physical Property Measurement System, Quantum Design) was used with a temperature range of 10 K – 363 K. Inside the PPMS, the vacuum environment is always a few ~$10^{-3}$ Torr, rendering convective heat losses negligible.

We sometimes found that the electrical resistance of the sensor slowly drifted (increased) with time at room temperature. However, this effect was stabilized after annealing the sample at 363 K for 5 min, eliminating resistance drift at room temperature. Therefore, this behavior could be related to the absorbed water on the metal electrodes. For the heater, we apply a sinusoidal voltage with frequency lower than 2 mHz through a standard resistor of 1 kΩ to flow current with a range of ±1.5 mA, generating sufficient heating power. To obtain the response of the sensor (thermometer), we measured its resistance change by a standard lock-in method with excitation frequency 2.147 kHz and current 1 μA (carefully chosen to avoid self-heating).

***Measurements and Data analysis*:** Figure S5a shows the measured sensor resistance change, $\Delta R_S$, as a function of the power applied to the heater, $P_H$, at $T$ = 100 K for the $SiO_2$ sample (Fig. S4). The black (for negative heater current, $I_H$) and red (for positive $I_H$) lines overlap with each other, indicating the measurement is symmetric and reliable. The calibration for sensor and heater resistance vs. temperature is shown in Fig. S5b; thus, sensor heating due to heater power $\Delta R_S$ can be converted to a temperature rise, $\Delta T_S$, as shown in Fig. S5c by using the resistance-temperature calibration curve. The fitted slope of the $\Delta T_S$ vs. $P_H$ curve in Fig. S5c is 0.01797 K/μW, which is then used for the extraction of thermal properties through simulations (see Section 3). Figure S5d shows the measured ratio of $P_H$ to $\Delta T_S$ for all representative samples as a function of ambient temperature from 20 K to 300 K. The uncertainty of the electrical thermometry measurement is ~2% (Tables 1 and 2), which is comparable to the symbol size on this plot. Thus, although data are available down to 20 K, the values are distinguishable without ambiguity only when $T \geq$ ~70 K, which is the temperature range shown in Figs. 2 and 4 of the main text.

We note that $P_H/\Delta T_S$ shown in Fig. S5d is not the thermal conductance through graphene, because $\Delta T_S$ is the temperature rise in the sensor, not the temperature drop from the heater to sensor, and $P_H$ is the heat generated in the heater, not the one flowing in graphene. The thermal conductance of the graphene cannot be immediately extracted from our raw data, due to heat leakage into the substrate (a drawback of the substrate-supported thermometry method). Instead, we employ 3D simulations to carefully account for all heat flow paths and, by comparison with the experiments, to obtain the thermal conductance of the graphene samples (see Section 3).

***Error analysis*:** Figure S6a shows the sensor resistance as a function of count number (time) without applying current to the heater at $T$ = 102 K. The standard deviation of the scattered data points is $\delta R$ = 3.1 mΩ, which corresponds to $\delta T$ ~ 36 mK by using calibration coefficient 0.0866

Ω/K obtained in Fig. S5b. Thus, the error of the temperature reading is ±36 mK, primarily due to slight ambient temperature fluctuations in the PPMS (consistent with a fluctuation of ±30 mK of the displayed temperature on the PPMS monitor). Zooming into the circled region of Fig. S6a, we note a resistance fluctuation $\delta R$ = 0.17 mΩ, corresponding to a temperature uncertainty ±2 mK due to electrical measurement instruments. Therefore, during the time scales of most of our measurements our temperature accuracy is limited by the ambient temperature control of the PPMS rather than by the electrical measurements themselves.

***Establishment of thermal steady-state*:** The sweep speed of the heater power is chosen to be sufficiently slow to reach thermal steady-state between the heater and sensor. Figure S6b shows the heater power ($P_H$) sweep with time and corresponding resistance change in the sensor, $\Delta R_S$. Data shown here correspond to the linear ramp in Fig. S5a. After ~15 minutes, the heater power reaches its maximum, and the change of sensor resistance follows the same trend without delay, indicating that the thermal steady-state between the heater and the sensor is established during the entire sweep process. If the sweep speed of the heater power is too fast to reach the steady-state, the data point at $P_H$ ~ 110 µW in Fig. S5a will deviate from the linear trend. We also verified this by a comparison between the corresponding constant DC power and the above methods.

## Supplementary Note 3: 3D Simulation to Extract Thermal Properties

To extract the thermal properties of graphene, GNRs, or the $SiO_2$ substrate from the measured $\Delta T_S$ vs. $P_H$, we use a commercial software package (COMSOL) to set up a three-dimensional (3D) finite element method (FEM) model of the entire structure. A typical setup is shown in Fig. S7, where only a half of the sample is included due to the symmetry plane which bisects the region of interest. The size of the Si substrate is 100×50×50 $\mu m^3$, covered by a 290 nm thick $SiO_2$ layer. While the simulated Si substrate is slightly smaller than the actual Si chip employed in practice (to manage computational complexity and meshing), the size of the simulated structure has been carefully chosen and verified to reproduce all heat flow through the substrate itself. Figure S7b shows the zoomed-in structure containing the core area of the thermometry, with GNRs, heater, and sensor highlighted by different colors. A more zoomed-in structure is shown in Fig. S7c, where from top to bottom different layers are 40 nm metal, 25 nm top oxide insulator, GNRs, 290 nm bottom oxide, and silicon, respectively.

To perform the simulation, the bottom surface and three side surfaces (except symmetry plane) of the Si substrate are held at the ambient temperature, *i.e.* isothermal boundary condition. Other outer surfaces of the whole structure are treated as insulated, *i.e.* adiabatic boundary condition. The Joule heating in the heater is simulated by applying a power density within the heater metal, and the stationary calculation is performed to obtain the temperature distribution in the steady state, as shown in Figs. 1e and S7d typically. After calculating the average temperature rise in the measured segment of the sensor, we obtain the simulated value of ($\Delta T_S/P_H$). Thus, the simulation effectively fits the thermal *conductance* ($G$) of the test sample between heater and thermometer. The thermal conductivity ($k = GL/A$) of the test sample is thus an effective fitting parameter within the FEM simulator, ultimately adjusted to yield the best agreement between the simulated and measured $\Delta T_S$ vs. $P_H$. This fitting process is implemented by using MATLAB to interface directly with the COMSOL software, taking ~0.5 hour on a single desktop computer to converge to a best-fit value at a single temperature point for a typical calculation.

Before performing a substantial amount of calculations, a series of optimizations were carried out. First, the mesh was optimized. Due to the extreme ratio of the graphene/GNRs thickness (~0.34 nm) to their typical in-plane dimensions (~10 μm), this subdomain was optimized using a swept mesh strategy rather than the typical free mesh. Other subdomains were optimized carefully using the free mesh strategy, and in the bottom oxide and Si substrate the mesh size grows gradually from the heating region to the boundaries. Second, the real substrate size is about 8×8×0.5 $mm^3$, which can be regarded as a semi-infinite substrate relative to the small heating region (~10 μm). In FEM modeling, however, we have to select a finite size for the substrate due to the computational limitation. By choosing the distance from the center of the heater to the side and bottom surfaces of the substrate as a testing variable, we found 50 μm is large enough to model this 3D heat spreading, consistent with the recent work by Jang *et al.*[5]. Third, the length of the six probe arms attached to the heater and sensor (see Fig. S7b) was chosen as 2 μm (shorter than their real counterparts), which was found to be sufficiently long to mimic any peripheral heat loss. Fourth, it was confirmed that the simulated $\Delta T_S/P_H$ is independent of the power $P_H$ applied in the heater, which means the final results do not rely on the choice of the power $P_H$.

## Supplementary Note 4: Thermal Properties of the $SiO_2$ Underlayer

To validate our thermometry approach, we have carefully focused on a control experiment to measure the thermal properties of the $SiO_2$ underlayer supporting our samples. The sample was first prepared as described before, including the graphene under the heater and thermometer electrodes, to reproduce the thermal contacts encountered in all samples. However, the exposed graphene was then etched by an oxygen plasma, leaving the bare $SiO_2$ as shown in Fig. S4 and Fig. 1b. Measurements were performed on part 1 and part 2 of the metal electrodes (see Fig. S4), and the analyzed sensor and heater temperature rises normalized by heater power as a function of the ambient temperature are shown in Fig. S8a.

To compare our 3D simulations to this experimental data set, we needed to fit the thermal properties of the $SiO_2$ layer (which dominate), and to a lesser degree those of the $SiO_2$-Si thermal boundary resistance (TBR) at the bottom of the $SiO_2$ layer (which also plays a role). While the thermal conductivity of $SiO_2$ is well-known and easy to calibrate against[44, 53], to the best of our knowledge no consistent data for the TBR of the $SiO_2$-Si interface ($R_{oxs}$) exist as a function of temperature. Two recent studies[54, 55] suggested $R_{oxs}$ ~ $5-7\times10^{-8}$ $m^2KW^{-1}$ at room temperature, but some earlier efforts[56-60] found the total TBR of metal-$SiO_2$-Si interfaces as low as ~$1-3\times10^{-8}$ $m^2KW^{-1}$, putting an upper bound on $R_{oxs}$ without being able to separate it from the total TBR. Due to this contradiction, we set out to obtain the temperature-dependent $R_{oxs}$, treating it as another fitting parameter of our simulations in addition to the thermal conductivity of $SiO_2$ ($k_{ox}$). Other thermal parameters well characterized in the literature are the thermal conductivity of highly doped Si[61-63], thermal boundary resistances of the graphene-$SiO_2$ interface[64, 65] and the Au-Ti-$SiO_2$ interfaces[4]. In addition, the effective thermal conductivity of the metal electrodes (Au/Ti) was calculated from the measured electrical resistance according to the Wiedemann-Franz Law, where an average Lorentz number $L$ = $2.7\times10^{-8}$ $W\Omega K^{-2}$ is used for Au/Ti electrodes[66]. (all parameters were allowed to vary within known experimental bounds, leading to the uncertainty analysis in Section 6 below.)

Our extracted $k_{ox}$ and $R_{oxs}$ of two data sets (part 1 and 2) are shown in Figs. S8b and S8c, respectively. Our $k_{ox}$ data are in a good agreement with well-established values reported by Cahill[44], and the typical uncertainty is ~5% at most temperatures. By fitting our $k_{ox}$ data with a polynomial up to the $7^{th}$-order, we obtained a smooth dependence of $k_{ox}$ on $T$ (green solid line), and this was used to extract the thermal properties of our graphene and GNRs. Our extractions suggest $R_{oxs}$ ~ $10^{-8}$ $m^2KW^{-1}$ at room temperature, in agreement with the reported upper bound in Refs. 56-60. For the subsequent thermal analysis, our $R_{oxs}$ data are best fit by a simple expression,

$$R_{oxs}(T) = \frac{1.046\times10^{-4}}{(T+13.4)^{2.25}} + 9.67\times10^{-9} \quad \mathrm{m^2K/W} \qquad \text{(S1)}$$

as shown by the green solid line in Fig. S8c. Thus, this control experiment demonstrates the feasibility and reliability of our thermometry platform, also giving the first report of the temperature-dependent TBR of $SiO_2$-Si interfaces.

## Supplementary Note 5: Additional Data and Comparisons

***All data of graphene samples:*** We examined three "short" graphene samples (GS1, 2, 3) which were not patterned into GNRs; all had length $L$ ~ 260 nm and width $W$ ~ 12 μm between the heater and thermometer electrodes. Besides GS1 which is shown in Fig. 1a of the main paper, the SEM image of another sample (GS2) is shown in Fig. S9a. The third one (GS3) broke after measurements at two temperature points, which are nevertheless listed among the data in Fig. S9. Figure S9b displays raw data taken as ratio of heater power to sensor temperature ($P_H/\Delta T_S$) for all three samples; the corresponding data of the $SiO_2$-only sample (part 2 of Fig. S8a) is also plotted here for comparison. The presence of graphene notably "heats" the sensor (higher $\Delta T_S$) and is distinguishable from the $SiO_2$-only sample all the way down to ~20 K (although the GNRs become harder to distinguish below ~70 K as mentioned earlier). The extracted graphene thermal conductivities are shown in Fig. S9c, and the three samples show very similar values. They all decrease from ~300 $Wm^{-1}K^{-1}$ at 300 K to ~10 $Wm^{-1}K^{-1}$ at 30 K and show similar temperature dependence. The data of the "long" graphene with $L$ ~10 μm from Ref. 6 are also plotted in Fig. S9c (black dots).

***Ballistic percentage:*** We compare the sample thermal *conductance* to the theoretical ballistic limits in Fig. S9d, recalling the relationship between conductance and conductivity, $G = kA/L$. The theoretical ballistic limit $G_{ball}/A$ is calculated by using the full phonon dispersion of graphene and integrating over the entire 2D Brillouin zone (see Section 9). Our three "short" graphene samples all display on average ~35% of the ballistic conductance limit, indicating they reach a quasi-ballistic conduction regime. The data for the 10-μm "long" sample from Ref. 6 show <2% of ballistic limit on average, indicating a diffusive transport regime as would be expected for a sample much longer than the phonon mean free path (mfp). Both percentages are consistent with a simple estimation of transmission probability $\sim\lambda_{bs}/(\lambda_{bs} + L)$ using their own lengths and back scattering mean free path $\lambda_{bs} = (\pi/2)\lambda \sim 160$ nm (see main text) where the intrinsic phonon mfp $\lambda \sim 100$ nm for most temperatures.

***Data with low-T range:*** Figure S10a shows the extracted thermal conductivity of our "short" graphene (GS1) and GNRs for the full temperature range measured, down to ~20 K (however, we recall that GNR measurements are challenging to distinguish below ~70 K, as previously mentioned). We can fit the thermal conductivity as $k = \alpha T^{\beta}$ below ~200 K, and the obtained power $\beta$ is shown as an inset of Fig. S10a. We find that $\beta$ decreases from ~1.6 for the unpatterned graphene to ~1 for narrow GNRs. However, we note that this does *not* necessarily mark a transition to one-dimensional (1-D) phonon flow, as the GNRs here are much wider than the phonon wavelengths (few nm). Thus, the simple model is given as a convenient analytic estimate, and the exponent $\beta$ represents the complex physical behavior of GNR heat flow due to the increasing heat capacity (which scales as $\sim T^{1.5}$) and the slightly decreasing phonon mfp in this $T$ range.

We also compare our extracted graphene and GNR thermal conductance with carbon nanotubes (CNTs) in Fig. S10b. We perform this comparison with the calculated ballistic upper limit, with a single-wall carbon nanotube (SWCNT) by M. Pettes *et al.*[45], and with multi-wall carbon nanotubes (MWCNTs) by P. Kim *et al.*[46], M. Pettes *et al.*[45], and M. Fujii *et al.*[47]. We find that our short graphene data (red filled dots) and P. Kim's MWCNT data[46] (open black squares) have the highest values, both reaching up to ~35% of the ballistic limit of graphene.

Figure S11 displays the width *W* dependence of our GNR thermal conductivity side-by-side with the electrical mobility measured by Yang and Murali[11] on similar samples (Fig. S11a is replot of Fig. 2d in the main text, here using log axis). Both the GNR thermal conductivity and electrical mobility show a similar trend with *W*, starting to decrease significantly when scattering becomes edge limited. However it is apparent that their fall-off occurs at different critical widths: *W* ~ 200 nm for thermal conductivity and *W* ~ 40 nm for electrical mobility. Above these widths, the thermal conductivity is limited by phonon-substrate scattering, while the electrical mobility is limited by electron scattering with substrate impurities. The difference between their critical *W* is consistent with the intrinsic phonon mfp $\lambda_{ph}$ ~ 100 nm being approximately five times larger than the intrinsic electron mfp[42] $\lambda_{el}$ ~ 20 nm in $SiO_2$-supported graphene. (We note phonons are entirely responsible for the thermal conductivity of these GNRs, with a negligible electronic contribution[13]). Thus, the fall-off of thermal conductivity and electrical mobility corresponds approximately to GNR widths approximately twice the phonon and electron mfps. Interestingly, this also suggests a GNR width regime (~40 < *W* < ~200 nm) where the thermal conductivity is reduced from intrinsic values but the electrical mobility is not yet affected by edge disorder. This suggests the possibility of manipulating heat and charge flow independently in such narrow edge-limited structures. Further control of such behavior could also be achieved if substrates with different roughness, impurity density, and vibrational (phonon) properties are used.

## Supplementary Note 6: Uncertainty Analysis

We can estimate the uncertainty of our analysis with the classical partial derivative method:

$$\frac{u_k}{k} = \sqrt{\sum_i \left( s_i \times \frac{u_{x_i}}{x_i} \right)^2} \tag{S2}$$

where $u_k$ is the total uncertainty in the extracted thermal conductivity $k$, $u_{xi}$ is the uncertainty in the $i$-th input parameter $x_i$, and the dimensionless sensitivities $s_i$ are defined by

$$s_i = \frac{x_i}{k}\frac{\partial k}{\partial x_i} = \frac{\partial (\ln k)}{\partial (\ln x_i)} \ . \tag{S3}$$

The partial derivatives are evaluated numerically by giving small perturbation of each parameter around its typical value and redoing the extraction simulation to obtain the change of $k$. To highlight the relative importance of each input parameter, we define their absolute contributions as $c_i = |s_i|\times(u_{xi}/x_i)$, and relative contributions as $c_i^2/\Sigma c_i^2$. The uncertainty analysis for extracting $k_{ox}$ and $R_{oxs}$ from the $SiO_2$ control experiment is performed in the same way.

Table S1 summarizes the sensitivities and uncertainty analysis for the unpatterned graphene (GS1) at 300 K. All input parameters can be separated into 3 classes: experimental data, thermal parameters, and geometric parameters. Their uncertainties are our estimates considering both random and systematic errors, and those of experimental and thermal parameters are updated appropriately as the temperature changes. The calculated sensitivities show that the graphene thermal conductivity is the most sensitive ($|s_i| > 2$) to the measured sensor response ($\Delta T_S/P_H$), thermal conductivity ($k_{ox}$) and thickness ($t_{box}$) of bottom $SiO_2$, center-to-center distances between metal lines of heater and sensor ($D_{met}$), between top $SiO_2$ lines of heater and sensor ($D_{tox}$), and between two voltage probes ($D_{pVV}$). These findings are consistent with previous work by W. Jang *et al.*[5] using similar substrate-supported thermometry structures. The input parameters with the greatest relative uncertainty ($u_{xi}/x_i$) are all three TBRs, thicknesses of top $SiO_2$ ($t_{tox}$) and metal ($t_{met}$), and the thermal conductivity of the Si substrate ($k_{Si}$) and metal ($k_{met}$). The combined effects of both sensitivities and relative uncertainties show that five largest contributions ($c_i > \approx 5\%$) to the overall uncertainty of our thermal measurement are from $\Delta T_S/P_H$, $k_{ox}$, $k_{Si}$, $R_{oxs}$, and $D_{met}$. In slight contrast to Ref. 5, we find that uncertainties introduced by $R_{oxs}$, $R_{mox}$ and $t_{met}$ are non-negligible for our structure and should be considered. On the other hand, geometric parameters related to the shape and size of the graphene sheet have very small sensitivities ($|s_i| < 0.001$), so their contributions are negligible in uncertainty analysis and not listed in Table S1.

For the extraction of GNR thermal conductivity, an example of calculated sensitivities and uncertainty analysis at 300 K is summarized in Table S2 (here for the sample with $W \sim 65$ nm). Compared with Table S1, we have two more parameters: the thermal conductivity of outer graphene connected to GNRs ($k_g$) and the width of GNRs ($W_{GNR}$). The parameters with the largest sensitivities ($|s_i| > 5$) are the same as those in Table S1, but their values increase because the total width of the GNR array is smaller than that of the unpatterned graphene. Due to the significant increase of sensitivities, the total uncertainty increases from 19% for unpatterned graphene to 60% for GNRs, while the input parameters with the greatest contributions ($c_i > 10\%$) to the total uncertainty are $\Delta T_S/P_H$, $k_{ox}$, $k_{Si}$, $R_{oxs}$, $D_{met}$, and $D_{tox}$, the same as those for graphene along with $D_{tox}$. From the 60% total uncertainty, approximately 21% is due to measurement uncertainty and the remainder from geometric and temperature-dependent variables as listed in Table S2. [we note that the geometric parameters related to the shape and size of GNRs and graphene *outside* the heater and sensor region have very small sensitivities ($|s_i| < 0.01$) and are not listed in Table S2, which is also consistent with the GNR results being insensitive to $k_g$.]

For other GNR samples with different widths, the total uncertainties gradually increase from 22% to 83% at 300 K as $W$ decreases from ~130 to ~45 nm, as less heat flows in the GNR array rather than the substrate. As the temperature decreases, the uncertainties of all graphene and GNR thermal conductivities also increase due to either increased sensitivities or increased relative uncertainties of input parameters.

As mentioned earlier and shown in Tables S1 and S2, all input parameters are classified into three groups, not all of which need to be included when comparing *relative* GNR thermal conductivities (e.g. with width or temperature). For instance, Fig. 2b in the main text compares the thermal conductivities of the samples at different temperatures considering the contributions of experimental and thermal parameters to the error bars. Similarly, Fig. 2d compares thermal conductivities of different samples at the same temperature, considering the contributions of experimental and geometric parameters to the error bars. Different samples share the same thermal parameters and these uncertainties would only shift all $k$ values in the same direction without affecting their relative values. In the end, the differences are relatively subtle, and within the fitting capabilities of our models, all based on the initial data above 70 K which are clearly distinguishable from one another as seen from the raw thermometry in Fig. S5d.

## Supplementary Note 7: Boltzmann Transport Equation (BTE) Model

The primary carriers of heat in graphene are phonons, at all temperature of interest in this study (20-300 K). In this section we develop a general model for thermal transport in graphene nanostructures based on the phonon Boltzmann transport equation (BTE). We consider scattering of phonons from 3-phonon interactions, line edge roughness, substrate roughness, $^{13}C$ isotopes, as well as corrections for the ballistic behavior in short GNRs. Due to the strong influence of edge disorder on transport in narrow GNRs, we employ a momentum-dependent model for the interactions of phonons with line edge roughness[31], while the remainder of scattering mechanisms is treated in the usual perturbation theory formalism.

In the direction of propagation, assuming a discrete distribution of phonon momentum states $\boldsymbol{q}$ along branches $\boldsymbol{s}$, the thermal conductivity can be expressed as

$$k(T) = \frac{1}{V}\sum_{s,q} \hbar\omega(q) v_s^2(q) \tau_{s,tot}(q) [\partial N(\omega,T)/\partial T] \qquad \text{(S4)}$$

with $v_s(q)$ the speed of sound in a branch $s$, $N = [\exp(\hbar\omega/kT)\text{-}1]^{-1}$ the equilibrium Bose-Einstein distribution function at temperature $T$ and energy $\hbar\omega$, and $\tau_{s,tot}(q)$ the total relaxation time.

First, it must be noted that thermal conductivity calculation calls for a volume $V$ of the sample (e.g. in $m^3$). In the case of 2D graphene, the volume is obtained by assuming a finite effective height of the monolayer graphene sheet, which we take to be[40] $H$ = 0.335 nm. This value reproduced experimental observations of the thermal conductivity of graphene flakes with excellent accuracy and will be used in the remainder of this section.

Second, in order to capture the dependence of the phonon velocity and energy on both the direction and magnitude of the phonon momentum, a full phonon dispersion relation is computed in the entire graphene Brillouin zone. We use a force constant (FC) method, as described by Saito *et al*.[67]. For this purpose, force constants are considered up to the fourth nearest neighbors. The force constants are taken from the improved model based on fitting the FC model to DFT results[67]. The full phonon dispersion relation and representation of the graphene Brillouin zone are shown in Fig. S12. We use the phonon dispersion of an isolated graphene sheet, which is a good approximation for $SiO_2$-supported graphene within the phonon frequencies that contribute most to transport[8], and at typical graphene-$SiO_2$ interaction strengths[7]. (However, we note that artificially increasing the graphene-$SiO_2$ coupling, e.g. by applying pressure[43], could lead to modifications of the phonon dispersion and hybridized graphene-$SiO_2$ modes[7].)

The total phonon scattering rate is a combination of isotopes/impurities, 3-phonon decays, edge roughness, substrate, and a ballistic correction term, such that

$$\tau_{tot}^{-1}(q) = \tau_I^{-1}(q) + \tau_U^{-1}(q) + \tau_{LER}^{-1}(q) + \tau_S^{-1}(q) + \tau_{ball}^{-1}(q)\,. \qquad \text{(S5)}$$

### 7.1 Impurity and Umklapp Scattering:

The semi-classical impurity scattering rate is given[31] by $\tau_I^{-1} = S_0\Gamma\rho(q)\omega^2(q)/4$. The strength of the interaction due to mass-difference scattering with isotopes (Γ**)** is calculated assuming the natural 1% concentration of $^{13}C$, with no other impurities assumed to be present, $S_0$ is the area of a cell of the first Brillouin zone (FBZ) and $\rho(q)$ is the vibrational density-of-states.

The model of 3-phonon decays distinguishes between of processes of type I (emission of a phonon) and type II (recombination of two phonons). Umklapp processes must satisfy conservation of momentum and energy. Using the set of destination nearest neighbor reciprocal cells (Fig. S12c), the decay of a phonon $q_0$ into $q_1$ and $q_2$ requires that

$$q_0 + b_i = q_1 + q_2$$
$$\omega_0(q) = \omega_1(q) + \omega_2(q) \qquad \text{(S6)}$$

Similarly, for type II umklapp processes, phonon absorption happens in accordance with conservation of energy and momentum

$$q_0 + q_1 = b_i + q_2$$
$$\omega_0(q) + \omega_1(q) = \omega_2(q) \qquad \text{(S7)}$$

Equations S6 and S7 above, with conditions on allowed combinations of phonon momentum vectors, define finite curves on the Brillouin plane. For instance, given an initial phonon momentum $q_0$ and branch $s_0$, the set of decay momentum $q_1$ from branch $s_1$ that allows decay with a branch $s_2$ is a line in the 2D first Brillouin zone. Thus, it is possible to approximate the delta function in order to realize energy conservation in the energy space. Computationally, handling the delta function requires particular care for its diverging nature. Here we use the following analytical representations of the Dirac function

$$\delta(\omega) = \lim_{\delta \to 0} \left( \frac{\delta}{\omega^2 + \delta^2} \right) \qquad \text{(S8)}$$

In computations, keeping the value of $\delta$ small but non-zero, gives an approximation to the Dirac function which respects the properties of symmetry and unicity of the function, and avoids introducing biases in computation of integrals. As a result, approximating $\delta(\omega_0(q_0) - \omega_1(q_1) - \omega_2(q_2))$ can be achieved by the following steps:

- For a given $q_0$ in the propagation direction, compute all allowed combinations of $q_1$, $q_2$ satisfying momentum conservation.
- For the allowed transitions, determined the corresponding phonon frequencies from the dispersion relation tables pre-computed using the FC method.
- Using the phonon frequencies above, compute delta functions for allowed transitions between bands.

Given the law of energy and momentum conservation, the 3-phonon scattering rate is computed from the following summation[40] for absorption

$$\tau_U^{-1}(s_0, q_0) = \frac{\hbar\Gamma^2}{3\pi\rho v_{s0}^2(q_0)} \sum_{1,2} \iint \omega_0(q_0)\omega_1(q_1)\omega_2(q_2)$$
$$\times [N(\omega_1) - N(\omega_2)]\delta(\omega_0(q_0) + \omega_1(q_1) - \omega_2(q_2))dq_1 \qquad \text{(S9)}$$

where $N$ is the equilibrium Bose-Einstein distribution, $\rho$ is the areal mass density of graphene, and $v_{s0}(q_0)$ is the group velocity of mode $q_0$ and branch $s_0$. Similarly for emission:

$$\tau_U^{-1}(s_0,q_0)=\frac{\hbar\Gamma^2}{3\pi\rho v_{S_0}^2(q_0)}\sum_{S_{1,2}}\iint \omega_0(q_0)\omega_1(q_1)\omega_2(q_2)$$
$$\times[N(\omega_1)+N(\omega_2)+1]\delta(\omega_0(q_0)-\omega_1(q_1)-\omega_2(q_2))dq_1 \qquad \text{(S10)}$$

where $\Gamma$ is the Gruneisen parameter. The value $\Gamma = 1$ used by Nika *et al.*[40] provides the best agreement with experimental results.

In Fig. S13e, the dependency of the umklapp scattering rate with $\omega^2$ clearly appears. Fig. S13f represents the computed total scattering rates in a short, $SiO_2$-supported GNR at $T = 300$ K. For acoustic branches, the model shows substrate and edge roughness scattering are the dominant scattering mechanisms in such a GNR. The optical modes are primarily limited by umklapp scattering, however they do not contribute to thermal conductivity due to very low occupancy at the temperatures of interest in this work. Given the results for total scattering rates in Fig. S13f, we can deduce that heat propagation in GNRs supported on $SiO_2$ mostly occurs through the longitudinal and transverse acoustic branches (LA and TA; also see Fig. 4 in the main text).

### 7.2 Edge Roughness Scattering:

In order to account for edge disorder, we compute a scattering rate between phonons and the line edge roughness (LER). Our model is based on partially specular interactions of phonons with LER and considers the interaction of lattice waves with the random variation at the GNR edges[31]. When a phonon wave reaches one of the edges, one of two things can happen: the phonon can either be reflected, resulting in a specular interaction which only flips the phonon momentum about the edge of the GNR without randomizing it, or it can be scattered diffusely, in which case the phonon momentum is randomized. A diffuse scattering event interrupts the flow of heat by randomizing the direction of the phonon propagation, thereby reducing thermal conductivity. The fraction of phonon interaction with the edge roughness that is specular is given by the specularity parameter $p(q)$, a number between 0 and 1. In order to capture the full dependence of LER interactions on the direction and magnitude of phonon momentum, as well as the root-mean-square (rms) amount of edge roughness ($\Delta$), we employ a momentum-dependent specularity parameter, given by

$$p(q)=\exp\left(-4q^2\Delta^2\sin^2\theta_E\right), \qquad \text{(S11)}$$

where $q$ is the magnitude of the phonon momentum, $\Delta$ is the rms roughness variation of the edges, and $\theta_E$ is the angle between the direction of phonon momentum and the edge. In this form, the specularity parameter is able to capture the details of the interaction between each phonon mode and the roughness at the edges[31].

However, the specularity parameter only gives us the probability of scattering each time a phonon interacts with an edge; in order to obtain the total rate of scattering with the edges, we have to trace each phonon mode through multiple possible specular interactions at the edges until it is either scattered diffusely at an edge, or scattered by another process while traveling between two opposite edges. Assuming phonons originate from locations which are uniformly distributed across the width of the ribbon, the average distance any phonon has to travel to the edge is $W/2$ along the direction normal to the edge. For a phonon mode $q$ and velocity vector $\boldsymbol{v}_s(q)$, the time it takes to cross the distance $W/2$ and reach the edge depends on the magnitude of the component of the phonon velocity vector in the cross-ribbon y-direction (along the direction normal to the edge, see Fig. S14). Then the velocity component in the cross-ribbon direction is $v_y(q)$ and the time it takes to reach the edge will be $W/[2|v_y(q)|]$.

Upon reaching the edge, the phonon has a probability [1-$p(q)$] of scattering diffusely at that edge, and a probability $p(q)$ of simply being reflected at the edge and continuing. In the latter case, the phonon then travels another distance of $W$ across the ribbon until reaching the opposite edge, where it can again scatter diffusely with a probability of [1-$p(q)$] or undergo another specular reflection and continue on its path. This process gives us an infinite series of terms accounting for all the possible scattering events and their probabilities as

$$\tau_{LER}(q) = \frac{W}{2|v_y(q)|}\left\{[1-p(q)] + 3p(q)[1-p(q)] + 5p^2(q)[1-p(q)] + 7p^3(q)[1-p(q)] + ...\right\}. \quad \text{(S12)}$$

This infinite series can be written as a product of two terms: a pre-factor $W/[2|v_y(q)|]\cdot[1-p(q)]$ and an infinite series in powers of $p(q)$ which can be summed in closed form as

$$\sum_{n=0}^{\infty}(2n+1)p^n(q) = \frac{1+p(q)}{[1-p(q)]^2} \quad \text{(S13)}$$

Then the total phonon lifetime due to scattering with the line edge roughness (LER) is given by

$$\tau_{LER}(q) = \frac{W}{2|v_y(q)|}\left[\frac{1+p(q)}{1-p(q)}\right]. \quad \text{(S14)}$$

We note here that this derivation[31, 68] unlike similar expressions previously derived for 3D structures[69], includes the dependence of the LER scattering rate on the angle between the phonon velocity direction and the edges through both the momentum- (and thus angle-) dependent specularity parameter $p(q)$ and the y-component of the phonon velocity vector.

### 7.3 Substrate Roughness Scattering:

The oxide on which the graphene is deposited allows for phonon recombination across the interface, and the presence of coupling between graphene and $SiO_2$ creates spatially variable perturbations of the Hamiltonian along the ribbon[6]. These exchanges can be modeled by considering a series of atomic points of contact between the graphene sheet and the substrate, due to the roughness of the oxide at the interface. The scattering rate on a phonon branch $i$ due to the combination of phonons leaking into the oxide, and geometrical perturbation of transport in graphene due to the rough contact points is[6]:

$$\tau_{S,i}^{-1}(q) = \frac{2\pi\Gamma}{4\omega_i^2}\left[\left|K_{fi}/\sqrt{M_C M_X}\right|^2 \rho_X(\omega_i) + \frac{\pi a^2}{S_0(1+1.5a\omega_i/v_i(q))}\left|K_{fi}/M_C\right|^2 \rho_C(\omega_i)\right] \quad \text{(S15)}$$

where $\Gamma$ is the proportion of scattering centers, $M_C$ and $M_X$ the molecular mass of carbon and oxide, and $\rho_C$ and $\rho_X$ the density of phonon states in graphene and oxide respectively. Here $K_{fi}$ are the interface force constants, whose values are obtained by Seol *et al.*[6]. In our interface model, $a$ is the average radius of a contact spot between the graphene and substrate due to $SiO_2$ roughness. We further assume a Gaussian distribution of asperities at the $SiO_2$ surface[70] with auto-covariance length $l_X$ and rms roughness $\Delta_{SiO2}$. In this assumption, asperities can be modeled by Gaussian shaped defects, whose radius of curvature at the summit is on average

$$R = \frac{l_X^2}{\Delta_{SiO_2}} \qquad \text{(S16)}$$

In modeling of the interaction between graphene and the substrate, since the graphene only partially conforms to the asperities of the rough substrate, we model hemispherical asperities on the $SiO_2$ surface to calculate the radius of the patches where the graphene is in contact with the substrate. We assume that the graphene is deposited on the asperity summits over a height $\delta$ = 0.5 Å such that the average radius of the spot contacts is

$$a = R\sqrt{1-(1-\delta / R)^2} \qquad \text{(S17)}$$

It is here remarkable that the second term of substrate scattering representing perturbations of heat conduction in graphene due to the substrate roughness varies mainly as ($l_X^4/\Delta_{SiO2}^2$). In particular, we note that an increase in substrate surface roughness rms results in *improved* conduction, as the graphene sample is in this case more “suspended”. Figure S13c represents the substrate scattering rates for acoustic branches that provide the best fit with the observed thermal conductivity in our collection of flakes and ribbons.

### 7.4 Ballistic Conduction Correction:

When the length of the GNR ($L$) becomes comparable to or smaller than the intrinsic phonon mean free paths (Figs. 4b,d in main text), then the influence of the contacts has to be taken into account. Since we treat the contacts as being large reservoirs, they can be assumed to be in equilibrium. This is also a good approximation in our experiments, where the contacts are large metal/oxide structures, approximately 200 nm wide and 70 nm tall, providing a large number of bulk vibrational modes to help equilibrate the phonons arriving from the GNR. The effect of contacts on phonon transport is to terminate the path of the GNR phonons, meaning that the longest phonon mean free path (mfp) cannot be longer than the physical length of the GNR (also see discussion on some small sub-continuum contact effects in the main text).

In the case of “short” GNRs, it becomes possible that a proportion of phonons transmits from one contact to the other without encountering other scattering events. In the extreme limit of complete absence of any other scattering, this behavior can be represented as ballistic transport under Landauer's formalism (see main text Fig. 2 and discussion); however, within our semi-classical BTE formalism it is convenient to combine the effect of the contacts with other scattering mechanisms. Therefore, we treat this effect as an additional corrective scattering mechanism that is included in our model, with lifetime $\tau_{ball}$, as derived below.

The lifetime of phonons due to the presence of contacts can be derived in a similar way to the treatment of line edge scattering. The effect of the contacts is similar to a completely diffuse boundary condition, as there is no reflection (the approximation of a perfectly absorbing contact) and the phonon momentum gets rapidly randomized when it enters a contact where the number of modes it can scatter with is very large. Then we consider phonons originating uniformly throughout the GNR and traveling toward one of the contacts with a velocity component $v_x(q)$ in the x-direction along the ribbon (see Fig. S14). The average time to reach the contact will then be $L/[2|v_x(q)|]$ for contacts separated by a distance $L$. Since there is no reflection at the contacts, the total phonon lifetime will simply equal the time to reach the contact, given by

$$\tau_{ball} = \frac{L}{2\left|v_x(q)\right|} \quad , \tag{S18}$$

which is consistent with the rate used for ballistic scattering by Lindsay *et al*.[71]. Since the phonon dispersion of graphene is not entirely isotropic, the result given in Eq. S18 is able to capture the angular dependence of the phonon velocity vector by taking the velocity component in the x-direction (along the GNR) as the relevant velocity for scattering due to contacts. The effect of the angular dependence of phonon velocity on the contact-limited ballistic scattering rate can be seen in Fig. S13a where we note that the rate depends not only on the energy of the phonons, but that different modes with the same energy can have different velocity components along the x-direction, leading to variation in the scattering rate.

## Supplementary Note 8: Additional BTE Results and Discussion

The frequency-dependent phonon mfp is directly obtained from the calculated total scattering rate (e.g. Fig. S13f). The frequency-dependent mfp is a weighted average for each branch as

$$\langle \lambda_s \rangle = \frac{\int c_s(\omega) v_s(\omega) \lambda_s(\omega) d\omega}{\int c_s(\omega) v_s(\omega) d\omega}, \tag{S19}$$

where $c_s(\omega)$ and $v_s(\omega)$ are the frequency-dependent heat capacity and group velocity, capturing the mfp with each frequency weighted by its contribution to the total thermal conductivity.

We show that the lattice thermal conductivity (Fig. S15a) and the phonon mfp (Fig. S15b) both scale with the length of the GNR in "short" ribbons due to the mfp being limited to about half the length, as expressed in Eq. S18 in the previous section. A difference exists between wide ($W$ = 2 μm) and narrow ($W$ = 65 nm) ribbons due to the presence of edge roughness scattering in the narrow ribbons. In the wide ribbons, the most significant scattering mechanism is substrate scattering, which limits the phonon mfp to around 100 nm (the value to which the total mfp converges for "long" ribbons in figure S15b). We note that this value is very nicely consistent with the average substrate-limited (or intrinsic) mfp deduced through a simple comparison with the ballistic conductance limits in the main text (Fig. 2 and surrounding discussion).

For a sample length $L$ ~ 200 nm, or approximately twice the intrinsic mfp, the length-dependent ballistic scattering rate is comparable to the substrate scattering rate and the effective thermal conductivity and phonon mfp become one-half of the substrate-limited values (Fig. S15a-b for "wide" sample). However, in narrow GNRs, the diffuse scattering at the edges limits the phonon mfp to approximately one-half of the width $W$ (the LER-limited mfp also depends on the rms edge roughness, as shown in Fig. S15d). Consequently, thermal transport in narrow GNRs is mostly diffusive until the length is reduced to values comparable to the width, and phonon transport again becomes partially ballistic (shown by solid lines in Fig. S15a-b).

We also observe that there is a limit to the effect of Δ on thermal transport: as edge roughness increases, more small-$q$ (large wavelength) phonons become scattered diffusely at the rough edges, as dictated by the specularity parameter in Eq. S11. However, strong substrate scattering in ribbons supported on $SiO_2$ also affects these long wavelength modes due to their lower energies (as can be deduced from the strong $1/\omega^2$ dependence of the substrate scattering rate in Eq. S15), leading to a saturation in the Δ dependence seen in Fig. S15d. Such separation of energy ranges where substrate and edge roughness scattering dominate opens the possibility of independent control of these two spectral components of thermal transport by adjusting substrate and edge roughness independently.

## Supplementary Note 9: Analytic and Empirical Modeling Notes

The theoretical ballistic limit $G_{\mathrm{ball}}/A$ (symbols in Fig. S16) is calculated by using the full phonon dispersion of graphene[19, 48] and integrating over the entire 2D Brillouin zone:

$$\frac{G_{ball}}{A} = \frac{1}{8\pi^2 H}\sum_s \int dq_x \int dq_y \hbar\omega_s(\vec{q})\,|\,v_{n,s}(\vec{q})\,|\,\frac{\partial N}{\partial T} \qquad \text{(S20)}$$

where $s$ is the phonon mode (branch), $H$ is the graphene thickness, $q$ is the phonon wavevector, $v_{n,s}$ is the phonon group velocity, $N$ is the equilibrium Bose-Einstein distribution. The obtained value at room temperature is $G_{\mathrm{ball}}/A$(300 K) = 4.16 GWK$^{-1}$m$^{-2}$; this value can differ by ±5% depending on the phonon dispersion used[40, 48, 67, 72], and is ~15% higher than obtained by only considering a simple 1D dispersion along the Γ-M direction[22]. (These observations were also pointed out previously in Ref. 19.) The numerically calculated $G_{\mathrm{ball}}/A$ can be well fitted by a simple analytical expression:

$$\frac{G_{\mathrm{ball}}}{A} \approx \left[\frac{1}{4.4\times10^5 T^{1.68}} + \frac{1}{1.2\times10^{10}}\right]^{-1} \qquad \text{(S21)}$$

over the temperature range 1-1000 K, as shown by the red line in Fig. S16. Below ~100 K, $G_{\mathrm{ball}}/A$ is dominated by the contribution of the flexural ZA modes, which scales as $\sim T^{1.5}$ due to its quadratic dispersion for low frequency phonons, $\omega \sim q^2$. The contributions of TA and LA modes scale as $\sim T^2$ due to their linear dispersion for low frequency phonons, $\omega \sim q$, and these become noticeable above 50 K. Thus, the combined $G_{\mathrm{ball}}/A$ scales with a power exponent that can be approximated as $\sim T^{1.68}$ over a wide temperature range (a little steeper than $\sim T^{1.5}$), as shown above.

The $L$-dependence of graphene thermal conductivity (solid lines in Fig. 2c) is obtained with the three-color model from Eq. 1 of the main text, where $G_{\mathrm{p,ball}}/A$ is taken from our calculation shown in Fig. S16, and $k_{\mathrm{p,diff}}$ is from theoretical simulations in Ref. 6, matching their experimental data. At $T$ = 300 K, $k_{\mathrm{diff}} \approx 560 \pm 50$ Wm$^{-1}$K$^{-1}$, and $k_{\mathrm{p,diff}}$ = 148, 214, 198 Wm$^{-1}$K$^{-1}$ for $p$ = ZA, TA, LA modes, respectively.

The $W$-dependence of GNR thermal conductivity (solid lines in Fig. 2d, the same as Fig. S11a) is obtained by Eq. 2 of the main text, where $k(L)$ is given by Eq. 1 [$k(L)$ = 346, 222, 158, 37 Wm$^{-1}$K$^{-1}$ for $T$ = 300, 190, 150, 70 K, respectively], $\Delta$ = 0.6 nm, $n$ = 1.8 for all displayed temperatures; here $c$ was fitted as 0.04019, 0.02263, 0.01689, and 0.00947 Wm$^{-1}$K$^{-1}$ for $T$ = 300, 190, 150, and 70 K, respectively, although $c$ cannot be assigned overly great physical meaning due to the limitations of the simple model, as explained in the main text.

## Supplementary References: